# Optimization of curved vortex generator geometry for enhanced thermal performance in water-to-air finned-tube heat exchanger

Josip Batista, Anica Trp*, Kristian Lenic, Mateo Kirincic

University of Rijeka, Faculty of Engineering, Vukovarska 58, 51000 Rijeka, Croatia

*Corresponding author, email address: anica.trp@riteh.uniri.hr

**Abstract**

In the study, the geometry of curved rectangular vortex generators (RVG) has been optimized to enhance heat transfer and overall performance of the analyzed water-to-air finned-tube heat exchanger (FTHE). The developed and validated conjugate numerical model includes a small section of the full-scale heat exchanger representing the computational domain, with fully developed velocity and temperature profiles specified at the inlet boundaries of the water. Four influential geometry parameters of the vortex generator were selected for the optimization; attack angle, curvature angle, axial position in the streamwise direction and height. The optimization process has been performed according to two criteria: maximization of heat transfer solely and maximization of heat transfer while considering the influence of pressure drop. Accordingly, one objective function related to the Colburn factor ($j$) and two objective functions combining the Colburn factor and the friction factor ($f$) were defined. The regression analysis has been used and the Central composite design of experiment was selected in order to obtain non-linear regression polynomial functions which were integrated into the investigated objectives. The optimization procedure has been performed using the genetic algorithm. Optimized configurations' performance improvement was evaluated through comparison with the performance of the reference FTHE with plain fins. For the RVG configuration optimized according to the objective of maximizing heat transfer, the Colburn factor was enhanced by 53.1%. Optimizing the RVG configuration to maximize heat transfer while accounting for the influence of pressure drop resulted in an enhancement of the FTHE performance factor $j/f$ by 14.4% for one set of geometry parameters, and an enhancement of $j/f^{0.33}$ by 26.5% for another.

## 1. Introduction

Finned-tube heat exchangers (hereafter, FTHE) are widely utilized in different engineering applications, including HVAC systems, power generation facilities, automotive cooling systems, aerospace technology and industrial processes. They are often used for heat exchange between two fluids, one in a gaseous phase (typically air) and the other in a liquid phase (typically water), where the convective heat transfer coefficient on the liquid side is significantly higher. Generally, between 70%–90% of the total thermal resistance is attributed to the convective thermal resistance on the gas side [1]. The fin geometry has been a subject of research for many decades. Over the time, various fin geometries have been developed and improved, with the most notable being plain fins, wavy fins, louvered fins, slotted fins and most recently fins with vortex generators (VGs).

The application of VGs belongs to the passive method of heat transfer enhancement. The VGs function like a special type of extended surface that can either be stamped onto or perforated from the fin [2]. Many researches have thoroughly investigated how the shape, size, position and orientation of VGs influence the performance of FTHEs. Various VG shapes were proposed in the literature, such as rectangular winglets, trapezoidal winglets, delta winglets and other, alternative types, e.g., envelope VGs [3], bow-shaped VGs [4] or arc-winglet VGs [5]. Wing-type VGs are less studied, since the winglet-type VGs are found to perform better by comparing the thermal performance in the rectangular air channel [6,7].

Apart from analyzing the effect of changes made to a specific geometry parameter of the winglet, optimization procedures were also carried out to discuss the combined effects of several parameters and find the optimal design concerning different objective functions. Jang et al. [8] conducted a numerical investigation to optimize span angle and transverse position of the rectangular block-type VG by using a conjugate-gradient method. The optimization aimed to maximize the area reduction ratio for the staggered and in-line FTHEs. Their findings revealed that the optimal design resulted in the area reduction ratio of 15–26% for the in-lined arrangement and 8–14% for the staggered arrangement. Salviano et al. [9] optimized longitudinal VGs in a two-row FTHE with inline and staggered tube configurations by using the SIMPLEX optimization procedure, while considering seven geometry parameters and two objective functions ($JF$ and $JF^{1/3}$). The numerical results have shown that improvement in heat transfer is more pronounced for staggered tube configuration. In addition, researchers concluded that the optimum VG shape is similar to the shape of rectangular winglet. Optimization of H-type FTHE with rectangular VGs and elliptical tubes through response surface methodology and genetic algorithm (GA) was performed by Tang et al. [10]. The optimum values of VG height, length, angle of attack and streamwise position improved performance index $(Nu/Nu_o)/(f/f_o)^{1/3}$ of heat exchanger by 48–55%. Ali et al. [11] used an adjoint-based optimization procedure to determine the optimal design of delta-winglet pairs (DWP), aiming to increase the Nusselt number while reducing friction losses in a rectangular channel. The optimized morphed geometry improved the overall performance by 8.6–10.3% compared to the case with flat DWPs. Using an artificial neural network (ANN) and GA, Xu et al. [12] optimized the attack angle, as well as the streamwise and transverse positions of X-shaped VGs. The $j/f$ factor of the optimized FTHE was improved by 110–130% compared to the reference case without VGs. The height of the wavy fins and attack angle of delta VGs were optimized by means of GA for the two objective functions, $j$ and $f$ factors, in the study conducted by Li et al. [13]. For the optimized wavy FTHE configuration, the factor $j$ was increased by 6% while also reducing the factor $f$ by 8% compared to the wavy fin geometry without VGs. Additionally, vortex generators can be designed with additional features, such as curved and wavy shapes [14] or punched holes [15], further improving the FTHE efficiency by reducing weight and pressure drop while maintaining or enhancing heat transfer performance. Some studies [16-18] have demonstrated the potential advantages of curved VGs over flat VGs in terms of the FTHE performance. Sarangi et al. [19] found optimum values of the curved trapezoidal VG geometry parameters (height, radius and subtended angle) that improve FTHE performance, i.e. minimize $f$ and maximize $j$ by using a combination of ANN and multi-objective GA (MOGA). Hu et al. [20] combined ANN and MOGA to optimize the transverse position of curved delta VGs inside the wavy FTHE, aiming to maximize heat transfer, while minimizing the required pumping power. For the optimized geometry, the heat transfer was increased by 35% under the same pumping power, compared to the baseline configuration without VGs. Sharma et al. [21] used the ANN and RSM data as surrogate models for GA optimization of the curved trapezoidal VG geometry parameters, with respect to the objective functions $j$ and $f$. The optimum values for VG heights, VG radius and VG angle formed at the tube center, which were obtained from the Pareto front, led to a significant improvement in FTHE performance. Sharma et al. [22] performed a similar optimization methodology with the same objective functions and optimized a novel type of combined flat and curved rectangular VGs.

Previous studies highlight the growing prominence of utilizing curved VGs to enhance air-side thermal performance in FTHEs. When optimizing geometry parameters of flat VGs, the attack angle is often considered as one of the most influential parameter. In contrast, the attack angle is neglected as an input variable for curved VG designs since the curved shape is often defined by describing a circular arc from the tube's central axis, thereby fixing the winglet's orientation relative to the flow. The optimization potential of curved winglet geometry parameters remains significant and offers considerable scope for further improvements.

In the present study, the geometry of curved rectangular VGs has been optimized to enhance thermo-hydraulic performance of the water-to-air FTHE under investigation. The conjugate numerical model that accounts for heat transfer on both fluid regions has been used based on the conclusions from [18, 23]. Four influential geometry parameters of the VG; attack angle ($\alpha_{\mathrm{RVG}}$), curvature angle ($\Theta_{\mathrm{RVG}}$), axial position in the airflow $x$ direction ($\Delta X_{\mathrm{RVG}}$) and height ($H_{\mathrm{RVG}}$), have been optimized according to two criteria; maximization of heat transfer only and maximization of heat transfer while considering the influence of pressure drop. These criteria are evaluated through three objective functions, enabling identification of different configurations that are optimal for either heat transfer enhancement or improved thermo-hydraulic performance. The applied RVG parametrization allows the winglet orientation to vary independently, thereby expanding the design space relative to conventional curved-VG definitions. The polynomial regression models have been defined to describe the relationship between the input variable parameters and the output responses. The approximated models have been integrated within the genetic algorithm. Results of the objective functions obtained using the single-objective and multi-objective GA approaches have been compared with numerical results for the reference FTHE with plain fins, used for the experimental validation.

## 2. Mathematical model of a physical problem

### *2.1. Problem description, computational domain*

The study focuses on fluid flow and heat transfer phenomena within the water-to-air FTHE. Hot water circulates through a staggered bundle of copper tubes, arranged in three rows, and transfers heat to the cold air which flows perpendicularly to the tube's axis through the gaps between thin aluminum fins (Fig. 1).

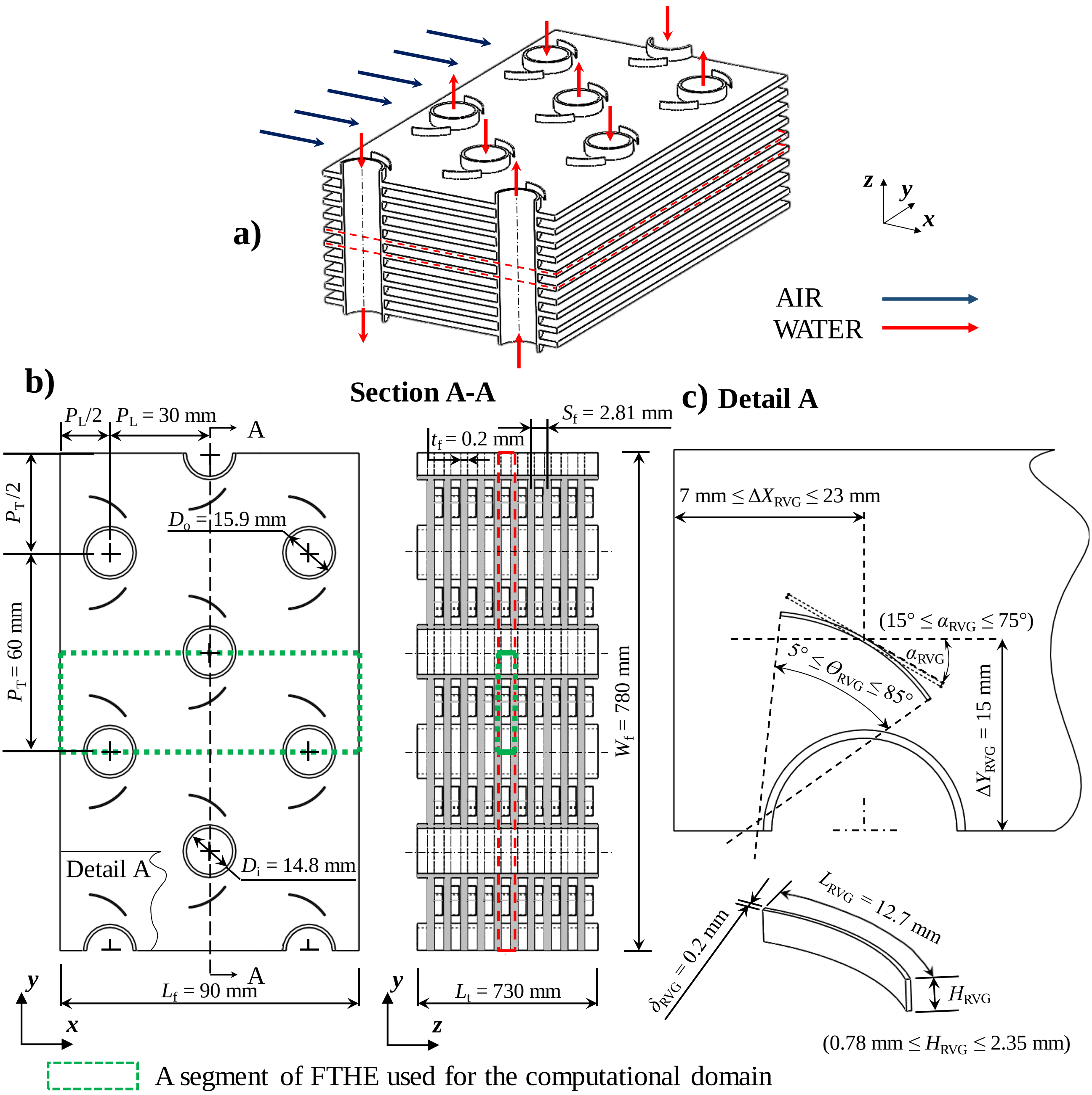

**Fig. 1.** Schematic 3D representation of the investigated FTHE (a), geometry parameters of fins and tubes (b), and RVGs (c).

For the analyzed segment shown in Fig. 1, the additional geometry characteristics are: minimum free-flow area $A_{min}$ = 57.6 mm$^2$, frontal area $A_{fr}$ = 84.3 mm$^2$, hydraulic diameter $D_h$ = 4.14 mm and heat transfer surface area density $\beta$ = 798 m$^2$/m$^3$. The curved rectangular winglets (RVG) have been symmetrically added to the plain fins. For the adequate description of the water-to-air heat transfer coupling problem, the proper approach would be to consider the whole FTHE. However, the computational domain can be simplified in order to reduce computational demands, without reducing the accuracy of the numerical solution significantly. Therefore, a small representative segment along the direction of the water flow is considered, for which the fully developed flow and thermal conditions (i.e., velocity and temperature profiles) are prescribed at the water inlets to capture the physical process of coupled heat transfer accurately enough. The computational domain, shown in Fig. 2, is divided into the following subdomains: air, water and wall, which includes fins, tubes and RVGs.

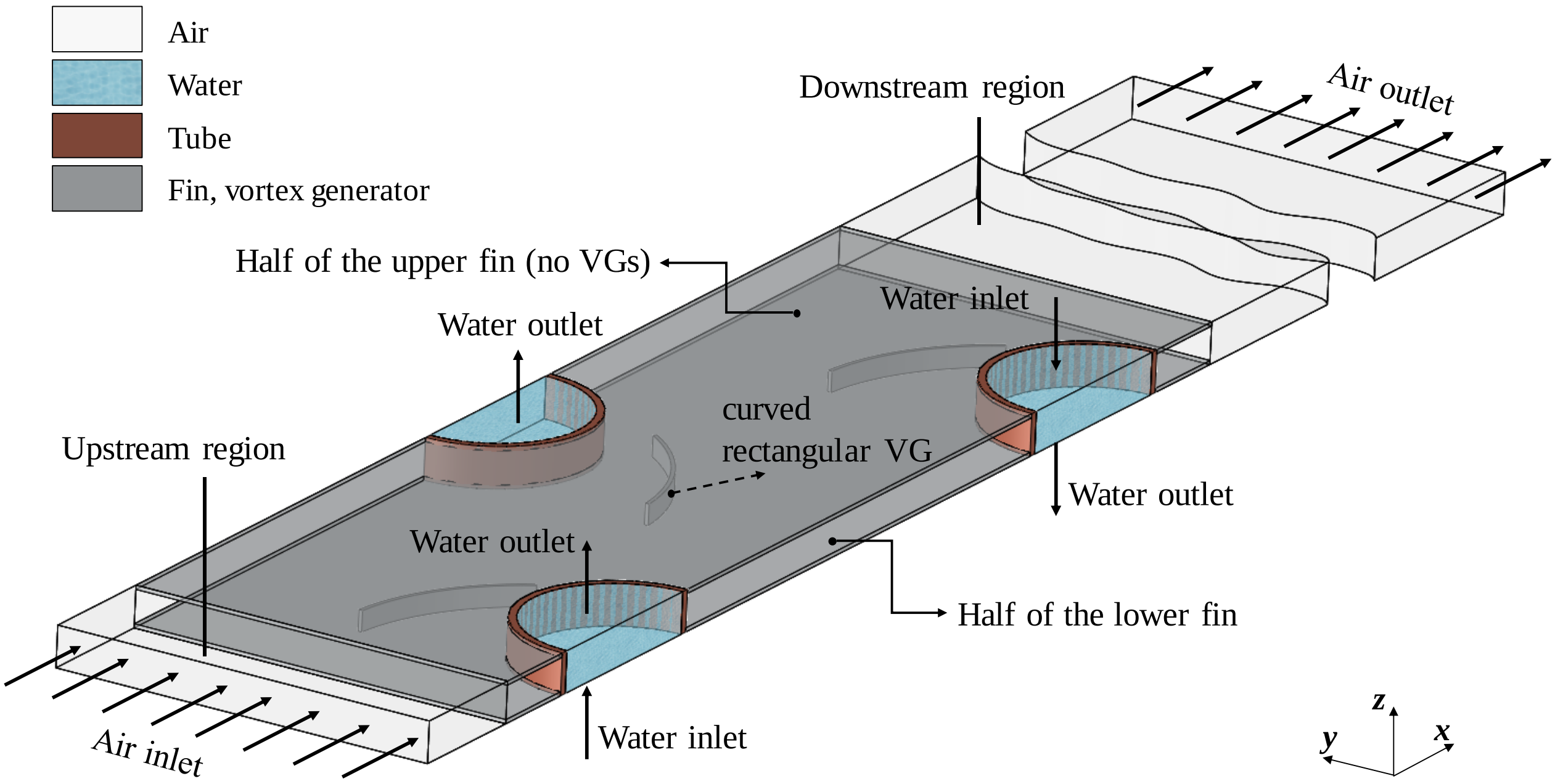

**Fig. 2.** The computational domain of the analyzed FTHE.

### *2.2. Conservation equations and boundary conditions*

The developed mathematical model defines a steady-state 3D mechanism of forced, steady, turbulent, incompressible fluid flow and heat transfer within a water-to-air FTHE. It is governed by a system of partial differential equations, along with appropriate boundary conditions. The fundamental governing equations for conservation laws [24] are described as follows.

Air and water subdomains

The continuity equation:

$$\frac{\partial w_i}{\partial x_i} = 0 \tag{1}$$

Momentum (2) and energy (3) equations:

$$\frac{\partial}{\partial x_i}\left(w_i w_j\right)=\frac{\partial}{\partial x_i}\left[\nu\left(\frac{\partial w_i}{\partial x_j}+\frac{\partial w_j}{\partial x_i}\right)-\overline{w_i^{'} w_j^{'}}\right]-\frac{1}{\rho}\frac{\partial p}{\partial x_j};\ i,\,j=1,\,2,\,3 \tag{2}$$

$$\frac{\partial}{\partial x_i}\left(w_i T\right)=\frac{\partial}{\partial x_i}\left[a\frac{\partial T}{\partial x_i}-\overline{T^{'} w_i^{'}}\right] \tag{3}$$

Here, terms $\overline{w_i' w_j'}$ and $\overline{T' w_i'}$ are Reynolds stresses and turbulent convective transport for constant density, respectively. The Boussinesq's hypothesis is used to approximate these terms:

$$\overline{w_i^{'} w_j^{'}}=-\nu_t\left(\frac{\partial w_i}{\partial x_j}+\frac{\partial w_j}{\partial x_i}\right)+\frac{2}{3}k\delta_{ij} \tag{4}$$

$$\overline{T^{'} w_i^{'}}=-a_t\left(\frac{\partial T}{\partial x_i}\right) \tag{5}$$

Due to its accuracy in near-wall regions, improved separation prediction, and robustness across different flow conditions, the two-equation SST *k-ω* turbulence model is used in this study [25]. The steady state conservation equations for turbulent kinetic energy *k* and specific dissipation rate *ω* are as follows:

$$\frac{\partial}{\partial x_i}\left(w_i k\right)=\frac{\partial}{\partial x_i}\left[\left(\nu+\frac{\nu_t}{Pr_k}\right)\frac{\partial k}{\partial x_i}\right]+G_k-D_k \tag{6}$$

$$\frac{\partial}{\partial x_i}\left(w_i \omega\right)=\frac{\partial}{\partial x_i}\left[\left(\nu+\frac{\nu_t}{Pr_\omega}\right)\frac{\partial \omega}{\partial x_i}\right]+G_\omega-D_\omega+Cd_\omega \tag{7}$$

where $Pr_k$ and $Pr_\omega$ are turbulent Prandtl numbers, $G_k$ and $G_\omega$ are generation terms, $D_k$ and $D_\omega$ are dissipation terms and $Cd_\omega$ is the cross-diffusion term. A detailed description of these coefficient can be found in [25, 26].

Solid subdomains (fins, tubes and RVGs)

The energy equation:

$$\frac{\partial}{\partial x_i}\left(\frac{\lambda_s}{c_s}\frac{\partial T}{\partial x_i}\right)=0 \tag{8}$$

To fully define the mathematical model of a given physical problem, it is necessary to specify the corresponding boundary conditions at the domain boundaries and the fluid-solid interfaces. The mathematical description of the boundary conditions is provided in Table 1.

**Table 1** The mathematical description of the boundary conditions.

| **Inlet** | **Outlet** | **Side boundaries parallel to *xz* plane** | **Top and bottom boundaries parallel to *xy* plane** | **Fluid-solid interface** |
|---|---|---|---|---|
| Air:<br>$w_x = w_{a,in}$<br>$w_y = 0$<br>$w_z = 0$<br>$T_a = T_{a,in}$ | Air:<br>$\partial w_x / \partial x = 0$<br>$\partial w_y / \partial x = 0$<br>$\partial w_z / \partial x = 0$<br>$\partial T_a / \partial x = 0$ | Upstream/downstream air region:<br>$w_y = 0$<br>$\partial w_x / \partial y = 0$<br>$\partial w_z / \partial y = 0$<br>$\partial T / \partial y = 0$ | Upstream/downstream air region:<br>$w_z = 0$<br>$\partial w_x / \partial z = 0$<br>$\partial w_y / \partial z = 0$<br>$\partial T / \partial z = 0$ | $\lambda_{solid} \frac{\partial T_{solid}}{\partial n} = \lambda_{fluid} \frac{\partial T_{fluid}}{\partial n}$<br>$w_n = 0$ |
| Water:<br>$w_x = w_x(x, y)$<br>$w_y = w_y(x, y)$<br>$w_z = w_z(x, y)$<br>$T_w = T_w(x, y)$ | Water:<br>$\partial w_x / \partial z = 0$<br>$\partial w_y / \partial z = 0$<br>$\partial w_z / \partial z = 0$<br>$\partial T_w / \partial z = 0$ | FTHE region:<br>$w_y = 0$ (fluid)<br>$\partial w_x / \partial y = 0$ (fluid)<br>$\partial w_z / \partial y = 0$ (fluid)<br>$\partial T / \partial y = 0$ (fluid, solid) | FTHE region:<br>$\partial T / \partial z = 0$ (solid) | |

*2.3. Definition of heat transfer and fluid flow parameters*

In this study, the tube outside diameter $D_o$ is used as the characteristic length when defining the air-side Reynolds number [13]. The air-side Reynolds number is determined as:

$$Re_a = \frac{w_m \cdot D_o}{\nu_a} \tag{9}$$

where $\nu_a$ is kinematic viscosity of the air and $w_m$ is the mean air velocity in the minimum free flow area. The corresponding water-side Reynolds number, calculated based on the inside tube diameter and inlet water velocity, was 17065.

The average heat transfer coefficient on the air side is calculated as:

$$h_a = \frac{1}{\eta_o \cdot A_o} \left( \frac{1}{U \cdot A_o} - \frac{\log_e \left( D_o / D_i \right)}{2 \cdot \pi \cdot \lambda_s \cdot L_t \cdot n_t} - \frac{1}{h_w \cdot A_{t,i}} \right)^{-1} \tag{10}$$

where $U$ is the overall heat transfer coefficient obtained from:

$$U = \frac{\dot{Q}}{A_o \cdot \Delta T_{lm}} \tag{11}$$

The water-side heat transfer coefficient $h_w$ has been evaluated using:

$$h_w = \frac{\dot{Q}}{A_{t,i} \left( T_b - T_{t,i} \right)} \tag{12}$$

where $A_{t,i}$ and $T_{t,i}$ are the inner tube wall surface area and the inner tube wall temperature, respectively, and $T_b$ is the bulk temperature on the water side.

In Eq. (10), the extended surface efficiency $\eta_o$ is defined as:

$$\eta_o = 1 - (1 - \eta_f) \frac{A_f}{A_o} \tag{13}$$

Here, $\eta_f$ denotes the fin efficiency and is calculated using an iterative approximation procedure proposed by Schmidt [27].

The Colburn factor ($j$) is a dimensionless parameter used to describe convective heat transfer on the air side:

$$j = \frac{Nu}{Re \cdot Pr^{1/3}} \tag{14}$$

where $Nu = h_a \cdot D_o / \lambda$ is the air-side Nusselt number and $Pr = \nu / a$ is the air-side Prandtl number.

The friction factor ($f$) quantifies the air-side pressure drop due to frictional resistance:

$$f = \frac{\Delta p_a}{0.5 \cdot \rho \cdot w_m \cdot c_p} \cdot \frac{A_{min}}{A_o} \tag{15}$$

## 3. Numerical procedure and model validation

### *3.1. Numerical solving*

The Ansys Fluent software is used for the numerical analyses of the fluid flow and heat transfer in the water-to-air FTHE. The SIMPLE algorithm is used to couple the pressure and velocity fields. The diffusion terms are discretized using second-order central-differencing scheme. For the discretization of the convective terms in the conservation equations, a second-order upwind scheme is selected. Convergence of the solution is assumed to be reached when the scaled residuals reduce below: $1 \cdot 10^{-6}$ for the continuity and momentum equations, $1 \cdot 10^{-5}$ for the $k$ and $\omega$ equations and $1 \cdot 10^{-9}$ for the energy equation.

A user-defined function is integrated into the standard calculation routine and coupled with the numerical solver to acquire the fully developed temperature and velocity profiles at the water inlet boundaries. The initial simulation was performed using uniform inlet profiles. After the convergence was obtained, the outlet velocity and temperature fields were applied as the inlet boundary conditions for the next simulation. To maintain the defined inlet temperature, the profiles were adjusted by the corresponding temperature drop along the flow path. The simulation was then repeated with updated profiles. Through the iterative procedure, fully developed profiles were obtained. A detailed description and validation of this method can be found in [23].

### *3.2. Meshing of the domain*

A structured hexahedral mesh is generated in the upstream and downstream air regions, and the remaining parts of the domain are discretized by a combination of structured and unstructured hexahedral meshes. A multi-zone meshing strategy was used, which automatically decomposes the geometry into blocks. In addition, the mesh is refined in the fluid regions near the solid walls to account for the high velocity and temperature gradients in the boundary layer formed. The mesh quality is assessed through orthogonality and skewness of the control volumes, and the respective average values of 0.99 and 0.037 are obtained.

A mesh independence study is carried out to ensure that the numerical results are not significantly influenced by the size or resolution of the computational mesh. In this study, the mesh is refined progressively to assess whether further refinements affect the solution. The numerical solutions for air outlet temperatures ($T_{a,out}$) and air-side pressure drops ($\Delta p_a$) are compared for four different mesh sizes: 0.74 million, 1.15 million, 1.62 million and 2.09 million control volumes and under the test conditions used in the optimization process (Fig. 3a). The error difference is below 1% for both $T_{a,out}$ and $\Delta p_a$ when the mesh size increases from 1.62 to 2.09 million control volumes. In addition, the near-wall mesh resolution was assessed using the area-averaged and maximum values of the non-dimensional wall distance ($y^+$). As shown in Fig. 3b, progressive mesh refinement reduced the corresponding values from 0.99 to 0.26 and from 10.43 to 2.55, respectively. For all meshes examined, the average $y^+$ remained below 1, indicating that the wall-adjacent region was resolved in accordance with the general $y^+ \approx 1$ requirement. Based on the results, the mesh size containing 1.62 million control volumes (Fig. 4) has been chosen.

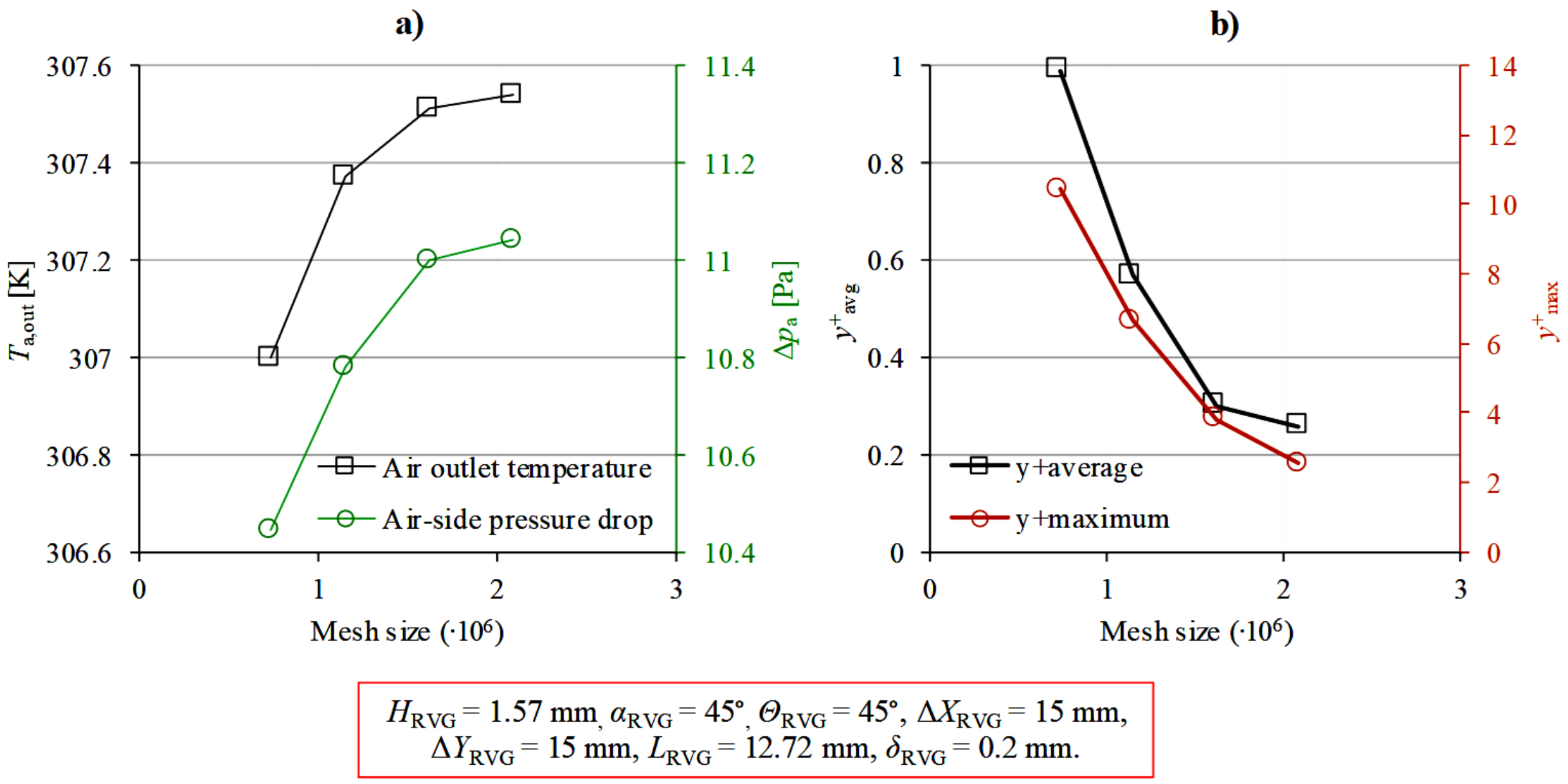

**Fig. 3.** Results of the mesh independence study: a) air outlet temperature and air-side pressure drop, and b) average and maximum non-dimensional wall distance ($y^+$).

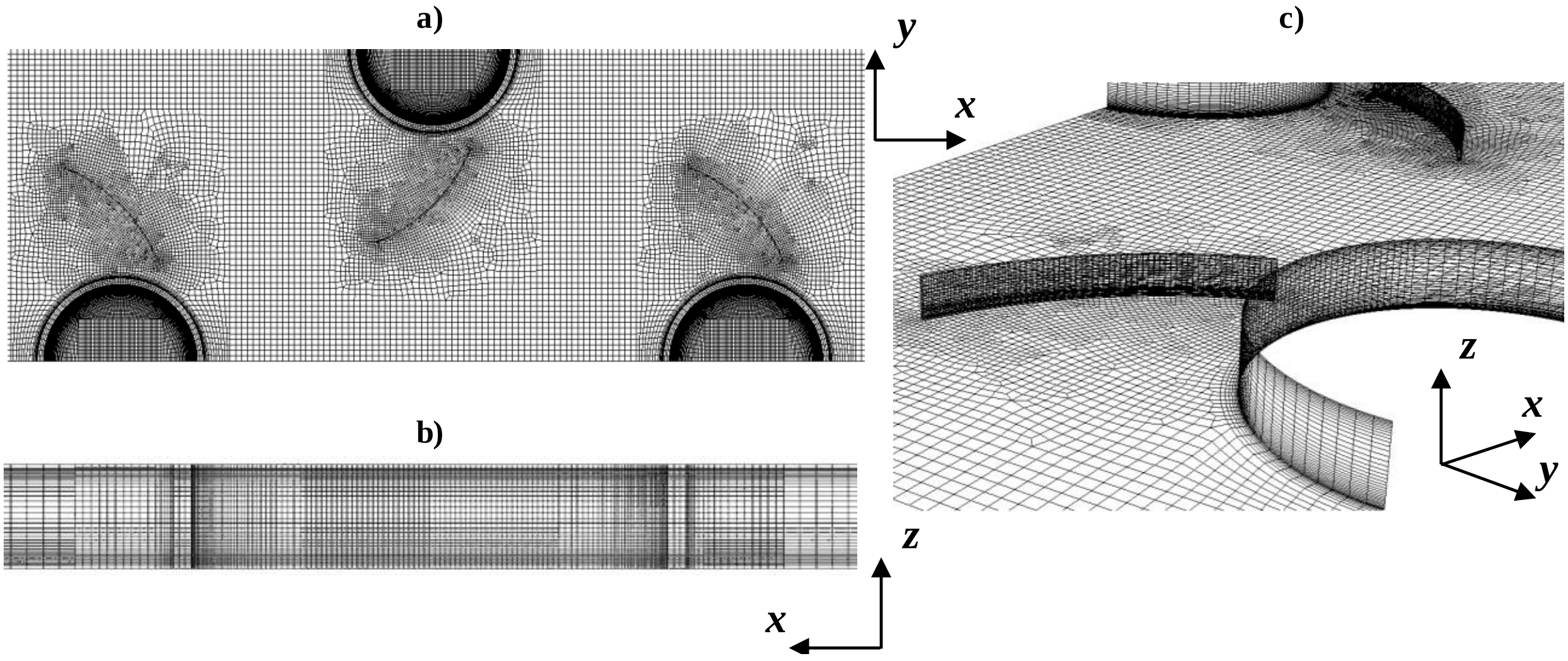

**Fig. 4.** Discretized computational domain: a) top view of the FTHE region, b) side view in the vicinity of the 2nd tube row, and c) local 3D view of the lower fin with VGs and tubes.

*3.3. Model validation*

The validity of the mathematical model and numerical procedure has been studied by comparison of numerical results with the measurement results obtained for water-to-air FTHE with plain fins (Fig. 5). The basic geometry parameters of the experimentally tested heat exchanger match those of the FTHE with curved RVGs. The experimental investigation was conducted in an open wind tunnel equipped with the FTHE test section, water circulation loop, and a data acquisition system. Air and water temperatures, airflow velocity, pressure drop, and water mass flow rate were measured using calibrated sensors integrated into the test facility. The air-side temperatures were measured by K-type thermocouples' arrays, whereas the water-side temperatures were measured by Pt100 resistance thermometers. Air temperature was measured at twenty measurement points. A measuring array consisting of five thermocouples was installed at 80 mm upstream of the test section, while the array of fifteen thermocouples was installed at 140 mm downstream of the test section. A total of four Pt100 sensors were used: two positioned upstream of the inlet header and two at the outlet of the distribution header. The air velocity at the inlet to the test section was measured using GrayWolf AS-202A hot-wire anemometers. The volumetric flow rate of water in the secondary flow circuit was measured using an Ultraflow 54 ultrasonic flow sensor connected to a Kamstrup Multical 403 multifunctional calorimeter. A Testo 521 differential pressure meter was used to measure the air-side pressure drop. Measurement data were collected and stored using a modular universal data acquisition system National Instruments SCXI 1000.

The measurement uncertainty of the calibrated thermocouples and PT100 sensors was ±0.35 °C and ±0.15 °C, respectively. The measurement uncertainty of the ultrasonic water flow sensor was ±2%. The measurement uncertainty of the differential pressure meter was ±0.5 Pa in the range from 0 to 20 Pa. The measurement uncertainty of the anemometers used for air velocity measurements was equal to the greater value of either ±0.015 m/s or ±2%.

The experimental tests on the heat exchanger were conducted under controlled and steady-state operating conditions on both the air and water sides. The first criterion for steady-state conditions requires that the calculated value of the average exchanged heat flux for each observed ten-second interval does not deviate by more than 5% from the average exchanged heat flux calculated at the end of the measurement. If the deviation exceeds this value, the measurement is excluded from further analysis due to excessive variability in the results. The second steady-state criterion requires that the difference between the exchanged heat flux on the water side and the air side, calculated at the end of the measurement using all available data, does not exceed 8%. A detailed overview of the experimental setup, data reduction methodology and uncertainty analysis can be found in [23]. The temperature and velocity ranges of air and water in the measurements served as the basis for the numerical simulations. The experiments were conducted under varying air-side operating conditions. The following operating parameter ranges were used: water inlet velocity was 0.76 m/s, water inlet temperature was 313 K, air inlet velocities were 0.455 m/s, 0.755 m/s and 1.03 m/s, and air inlet temperature was 295 K.

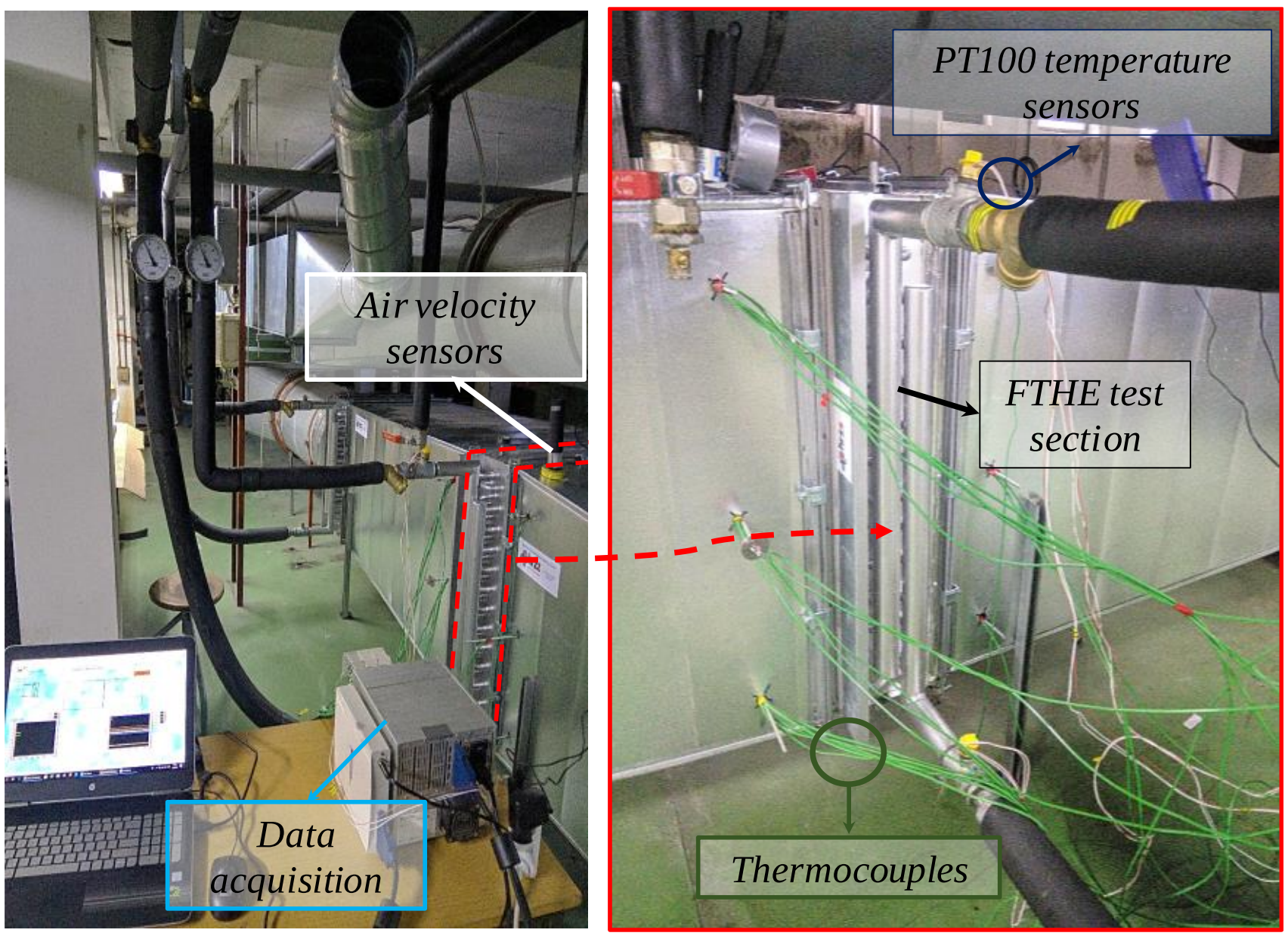

**Fig. 5.** The water-to-air FTHE test section with embedded temperature and velocity sensors.

Fig. 6 shows a comparison between experimental and numerical results of average air temperatures at FTHE outlet ($T_{a,out}$) and pressure drops on the air side ($\Delta p_a$) . The maximum difference in $T_{a,out}$ across the tested range of air-side operating conditions is 0.29 °C for $w_{a,in}$ = 0.755 m/s. Also, good agreement is observed for $\Delta p_a$, with relative deviations of 12.85% for $w_{a,in}$ = 0.455 m/s and 7.86% for $w_{a,in}$ = 1.03 m/s. The differences fall within the accuracy range of thermocouples and differential pressure measurement device.

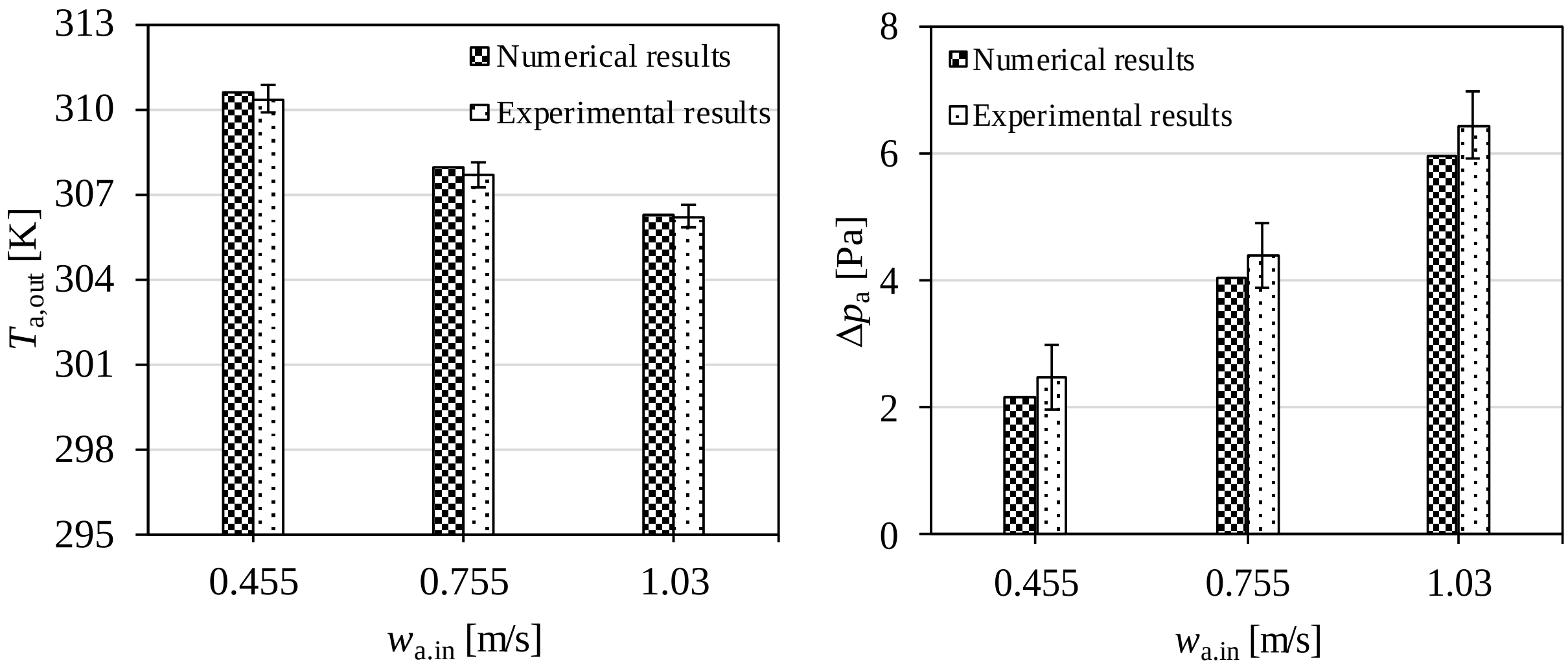

**Fig. 6.** Validation of numerical results against experimental data: average air temperatures at FTHE outlet (left) and air-side pressure drops (right), for test conditions: $T_{a,in}$ = 295 K, $T_{w,in}$ = 313 K, $w_{w,in}$ = 0.76 m/s.

Mathematical model and numerical procedure are further validated against the correlation of air-side Nusselt number, proposed by Xie et al. [28] for FTHEs with plain fins and large-diameter tubes. The prediction accuracy of correlation is ±10% in the Reynolds number range of 1000 < $Re_a$ < 6000. The comparison is given for air inlet velocities of 1.03 m/s, 1.5 m/s and 2 m/s which resulted in $Re_a$ values of 1524, 2220 and 2960, respectively. The comparison of $Nu$ vs. $Re_a$ is shown in Fig. 7. Relative to the correlation [28], the present numerical model underestimated the Nusselt number only by 0.9%, 2.3% and 4.6% for $Re_a$ = 1524, 2220 and 2960, respectively. The observed agreements suggest that the mathematical model, as well as the corresponding numerical procedure, can accurately describe the physical problem of heat transfer and fluid flow in the water-to-air FTHE and can therefore be used for further optimization of the investigated FTHE.

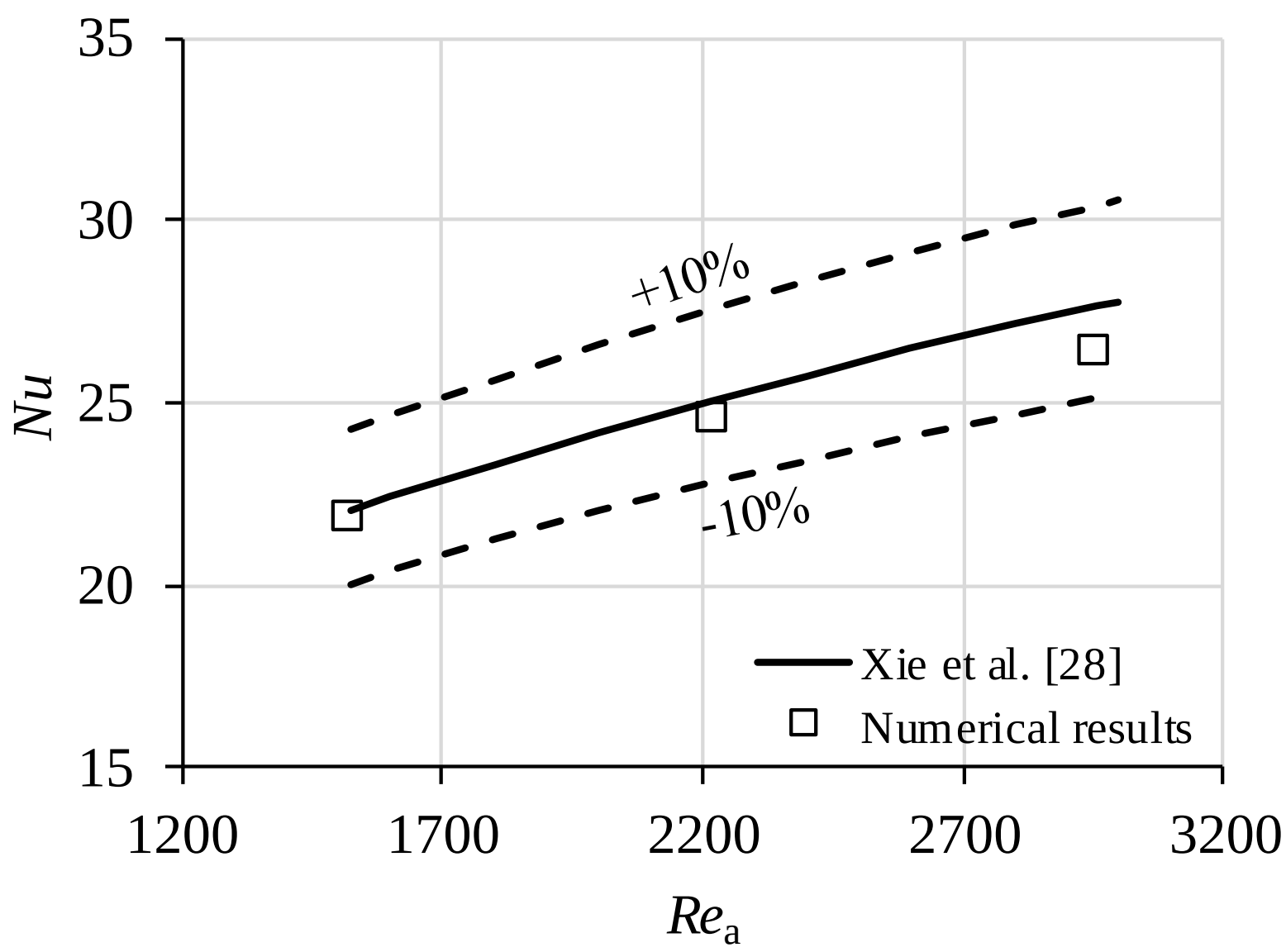

**Fig. 7.** The comparison between the numerically determined air-side Nusselt number and the literature correlation.

## 4. Optimization procedure

### *4.1. Selection of geometry parameters and objective functions*

For the purpose of optimization, the following variable RVG geometry parameters with the significant impact on the thermal and flow characteristics within the analyzed FTHE are selected, along with the corresponding ranges of their values:

- winglet attack angle, $15° \leq \alpha_{\mathrm{RVG}} \leq 75°$,
- winglet curvature angle, $5° \leq \Theta_{\mathrm{RVG}} \leq 85°$,
- winglet axial position in the airflow *x* direction, $7\ \mathrm{mm} \leq \Delta X_{\mathrm{RVG}} \leq 23\ \mathrm{mm}$,
- height of the winglet, $0.78\ \mathrm{mm} \leq H_{\mathrm{RVG}} \leq 2.35\ \mathrm{mm}$.

The values of RVG geometry parameters, as well as the values of inlet velocities and temperatures which were constant throughout the optimization procedure are shown in Table 2. Tube and fin dimensions of the analyzed heat exchanger with vortex generators corresponded with the reference FTHE with plain fins used for experimental validation. The operating conditions considered in the optimization study were chosen in accordance with those applied during the measurements.

**Table 2** Air and water inlet conditions and constant RVG geometry parameters used in the optimization procedure.

| Dimensions and operating parameters | Value |
|---|---|
| Winglet axial position from the tube center in the *y* direction, $\Delta Y_{\mathrm{RVG}}$ | 15 mm |
| Length of the winglet, $L_{\mathrm{RVG}}$ | 12.72 mm |
| Thickness of the winglet, $\delta_{\mathrm{RVG}}$ | 0.2 mm |
| Air inlet temperature, $T_{\mathrm{a,in}}$ | 295 K |
| Water inlet temperature, $T_{\mathrm{w,in}}$ | 313 K |
| Air inlet velocity, $w_{\mathrm{a,in}}$ | 1.03 m/s |
| Water inlet velocity, $w_{\mathrm{w,in}}$ | 0.76 m/s |

Optimum set of values for the analyzed geometry parameters is determined with respect to defined criteria. The first criterion is aimed at the enhancement of the dimensionless air-side convective heat transfer coefficient, i.e. the Colburn factor. The second criterion represents the demand for achieving the most favorable thermo-hydraulic characteristics, which simultaneously take into account both the heat transfer and the air-side pressure drop. Combined flow and heat transfer characteristics are determined by the $JF_{\mathrm{I}}$ and $JF_{\mathrm{II}}$ factors. Furthermore, for the purpose of optimization, two independent functions $f_1(x_i)$ and $f_2(x_i)$ are introduced, through which the three objective functions $F_1$, $F_2$ and $F_3$ are evaluated. The first function $f_1$ is defined as the ratio of the Colburn factors of the FTHE with rectangular VGs ($j_{\mathrm{RVG}}$) and the reference heat exchanger with plain fins ($j_{\mathrm{ref}}$), which can be written as:

$$f_1\left(\alpha_{\mathrm{RVG}}, \Theta_{\mathrm{RVG}}, \Delta X_{\mathrm{RVG}}, H_{\mathrm{RVG}}\right) = j_{\mathrm{RVG}} / j_{\mathrm{ref}} \tag{16}$$

The second function $f_2$ is defined as the ratio of the friction factors of the FTHE with rectangular VGs ($f_{RVG}$) and the reference heat exchanger with plain fins ($f_{ref}$), i.e.:

$$f_2\left(\alpha_{RVG}, \Theta_{RVG}, \Delta X_{RVG}, H_{RVG}\right) = f_{RVG} / f_{ref} \tag{17}$$

The objective function $F_1$, relating to the demand for the enhancement of heat transfer, is equal to the function $f_1$, as:

$$F_1 = f_1 = j_{RVG} / j_{ref} \tag{18}$$

The objective functions $F_2$ and $F_3$, relating to the demand for the enhancement of heat transfer, while taking into account the influence of pressure drop, are defined as follows:

$$F_2 = f_1 / f_2 = \frac{j_{RVG} / j_{ref}}{f_{RVG} / f_{ref}} = JF_{I} \tag{19}$$

$$F_3 = f_1 / f_2^{0.33} = \frac{j_{RVG} / j_{ref}}{\left(f_{RVG} / f_{ref}\right)^{0.33}} = JF_{II} \tag{20}$$

The objective function $F_2$ reflects the trade-off between the enhancement of heat transfer and the increase in pressure drop at a constant flow rate, while the objective function $F_3$ accounts for both heat transfer enhancement and pressure drop but under the constraint of maintaining the same pumping power [29].

*4.2. Regression modelling and optimization algorithm selection*

The central composite design (CCD) [30] used in the present study is characterized by a five-level variation of factors $k$. In this design, not all possible combinations of analyzed parameters are examined; rather, the examined points are distributed according to:

- two-level full factorial design ($2^k$),
- central state, where each factor has a central value,
- axial state $a$, composed of $2k$ axial points equally positioned from the central point.

The selection of the number of central points and the values of axial state defines the characteristics of the CCD model, i.e., its rotatability and orthogonality. The constraints were introduced so that the values of selected geometry parameters are coded at five levels: lower axial ($-a$), lower factorial (-1), central (0), upper factorial (+1) and upper axial ($+a$). The axial points are positioned at distances ±2 from the central point so that the adjustable parameters can be varied on a total of five levels.

It is necessary to select the suitable number of test points so adequate precision with acceptable operation time is retained. Compared with the three level fully factorial design, which requires 81 combinations for four parameters, the CCD for the present optimization problem requires 31 combinations of analyzed parameters (of which 25 are different). Since approximately 24 h of computational time was required for a single simulation to achieve the fully developed boundary condition at water inlets, the total number of design points had to be carefully limited in order to maintain a reasonable computational cost. From this perspective, the use of the CCD approach was justified. For each combination, a geometry is designed and a corresponding mathematical model is defined, which is then numerically solved. Table 3 shows the range of values for each of the four selected parameters.

**Table 3** Values of analyzed geometry parameters used in optimization process.

| Geometry parameter | Coded/real values | | | | |
|---|---|---|---|---|---|
| | -$a$ (-2) | -1 | 0 | +1 | +$a$ (+2) |
| RVG attack angle, $\alpha_{RVG}$ (°) | 15 | 30 | 45 | 60 | 75 |
| RVG curvature angle, $\Theta_{RVG}$ (°) | 5 | 25 | 45 | 65 | 85 |
| Axial position in the airflow $x$ direction, $\Delta X_{RVG}$ (mm) | 7 | 11 | 15 | 19 | 23 |
| RVG height, $H_{RVG}$ (mm) | 0.78 | 1.17 | 1.57 | 1.96 | 2.35 |

Using the regression analysis and the CCD, polynomial equations for the functions $f_1$ and $f_2$ are obtained, which describe the relation between the input and output variables. Regression equations are as follows:

$$\begin{aligned} f_1 = j_{RVG} / j_{ref} &= 0{,}424 + 4.46\times10^{-3}\alpha_{RVG} + 2.81\times10^{-3}\Theta_{RVG} + 1.268\times10^{-2}\Delta X_{RVG} + 0.7739 H_{RVG} - \\ & 8.8\times10^{-5}\alpha_{RVG}^2 - 1.2\times10^{-5}\Theta_{RVG}^2 - 1.312\times10^{-3}\Delta X_{RVG}^2 - 0.241\cdot H_{RVG}^2 - 2.8\times10^{-5}\alpha_{RVG}\Theta_{RVG} + \\ & 3.31\times10^{-4}\alpha_{RVG}\Delta X_{RVG} + 1.662\times10^{-3}\alpha_{RVG}H_{RVG} + 4.8\times10^{-3}\Delta X_{RVG}H_{RVG} \end{aligned} \quad (21)$$

$$\begin{aligned} f_2 = f_{RVG} / f_{ref} &= 1{,}02 - 7.32\times10^{-2}\alpha_{RVG} + 6.46\times10^{-4}\Theta_{RVG} + 4.36\times10^{-2}\Delta X_{RVG} + 1.92 H_{RVG} + \\ & 1.89\times10^{-4}\alpha_{RVG}^2 - 9\times10^{-4}\Delta X_{RVG}^2 - 2.432 H_{RVG}^2 + 1.26\times10^{-3}\alpha_{RVG}\Delta X_{RVG} + 6.93\times10^{-2}\alpha_{RVG}H_{RVG} - \\ & 1.14\times10^{-2}\Delta X_{RVG}H_{RVG} + 0.681 H_{RVG}^3 - 1.776\times10^{-3}\alpha_{RVG}\Delta X_{RVG}H_{RVG} \end{aligned} \quad (22)$$

The significance level is set at 0.05. The obtained $p$-values, as well as the coefficients of determination $R^2$ and $R_{adj}^2$ for the proposed regression models in expressions (21) and (22), are shown in Table 4. Since the values of $R^2$ and $R_{adj}^2$ are above 90% for both models, it can be concluded that the polynomial functions obtained through regression accurately represent the numerically obtained responses. The $p$-value for both regression models is 0, indicating that the relationship between independent variables and dependent variable is statistically significant, with negligible chance that the observed results occurred by randomness.

**Table 4** Validation of regression models based on the coefficients of determination and $p$-value.

| **Regression models** | | $R^2$ | $R_{adj}^2$ | **$p$-value** |
|---|---|---|---|---|
| $f_1 = j_{RVG}/j_{ref}$ | (21) | 97.54% | 95.90% | 0 |
| $f_2 = f_{RVG}/f_{ref}$ | (22) | 99.36% | 98.93% | 0 |

The Pareto charts in Fig. 8 summarize the standardized effects of the investigated geometry parameters and their interaction terms on thermo-hydraulic performance indicators, with Fig. 8a corresponding to $f_1 = j/j_{ref}$ and Fig. 8b to $f_2 = f/f_{ref}$. The red vertical line denotes the 95% confidence limit, while the bar length represents the relative magnitude of each effect. In addition, the signs '+' or '–' indicate whether the corresponding term is associated with an increase or a decrease in the response.

For $f_1 = j/j_{\text{ref}}$, the largest contributions are associated with $H_{\text{RVG}}$ and $\alpha_{\text{RVG}}$, and both terms show a positive influence on the response. Several quadratic terms, namely $(H_{\text{RVG}})^2$, $(\Delta X_{\text{RVG}})^2$ and $(\alpha_{\text{RVG}})^2$, also contribute noticeably and indicate a decreasing trend in $j/j_{\text{ref}}$. In addition, the interaction term $\alpha_{\text{RVG}} \times \Delta X_{\text{RVG}}$ shows a further positive contribution. For $f_2 = f/f_{\text{ref}}$, the dominant effect is related to $\alpha_{\text{RVG}}$, followed by $\Delta X_{\text{RVG}}$, $H_{\text{RVG}}$ and $\alpha_{\text{RVG}} \times H_{\text{RVG}}$. The terms $\alpha_{\text{RVG}}$ and $H_{\text{RVG}}$ are associated with an increase in $f/f_{\text{ref}}$, whereas $\Delta X_{\text{RVG}}$ and some related interaction terms are associated with a decrease in the response.

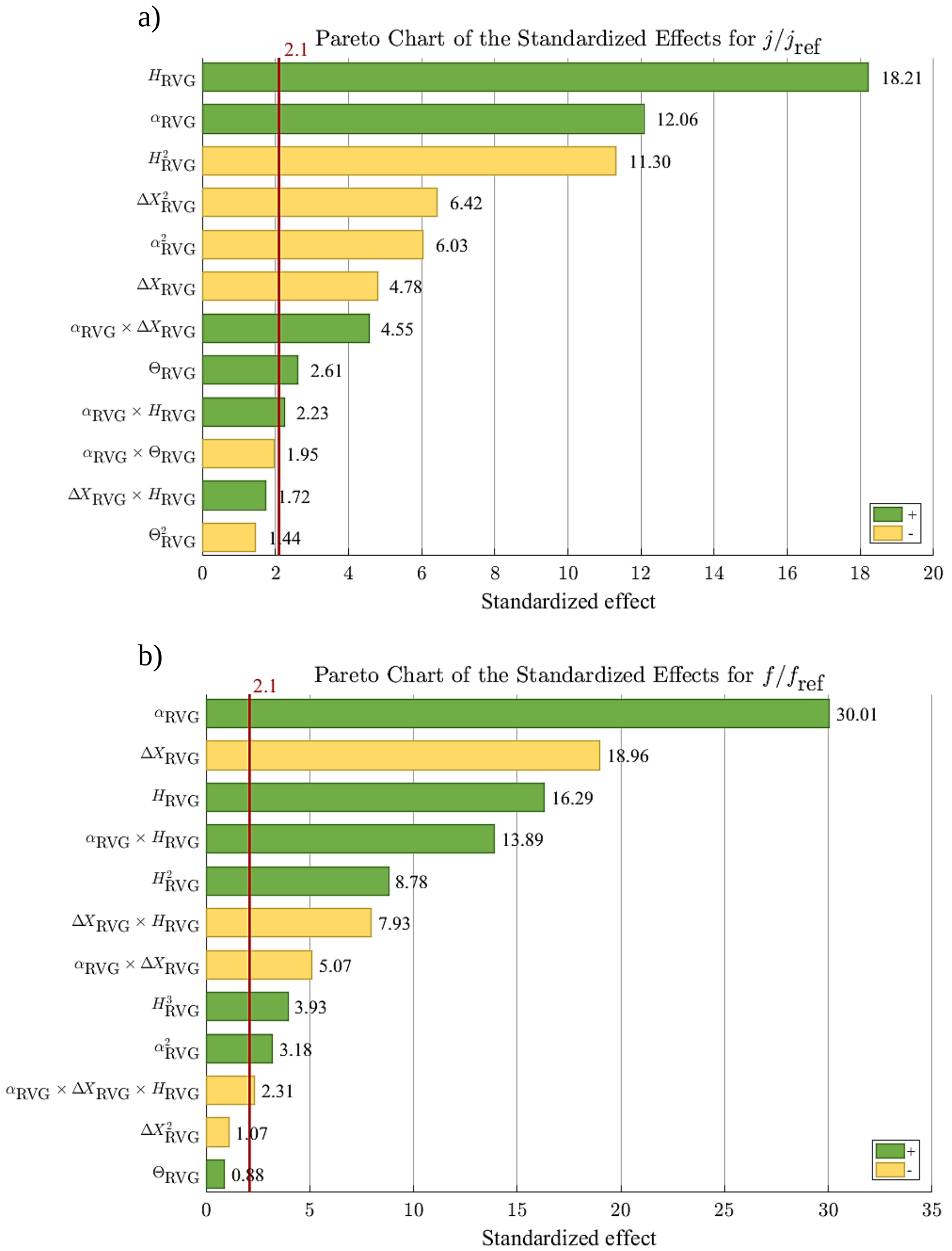

**Fig. 8.** Pareto chart of the standardized effects for: a) $f_1 = j/j_{\text{ref}}$ and b) $f_2 = f/f_{\text{ref}}$.

The above regression models can be used to predict various combinations of design variables. Fig. 9 compares the numerically obtained and estimated values of functions $f_1$ and $f_2$, where the subscript 'pr' represents the estimated values using Eq. (21) and Eq. (22), and the subscript 'num' represents the numerically obtained values. The observed deviations are within ±3% for $f_1$ and within ±8% for $f_2$, suggesting that the fitted correlations are reasonably accurate and can be used to predict thermo-hydraulic behavior of the FTHE with differently configured RVGs, under given constraints.

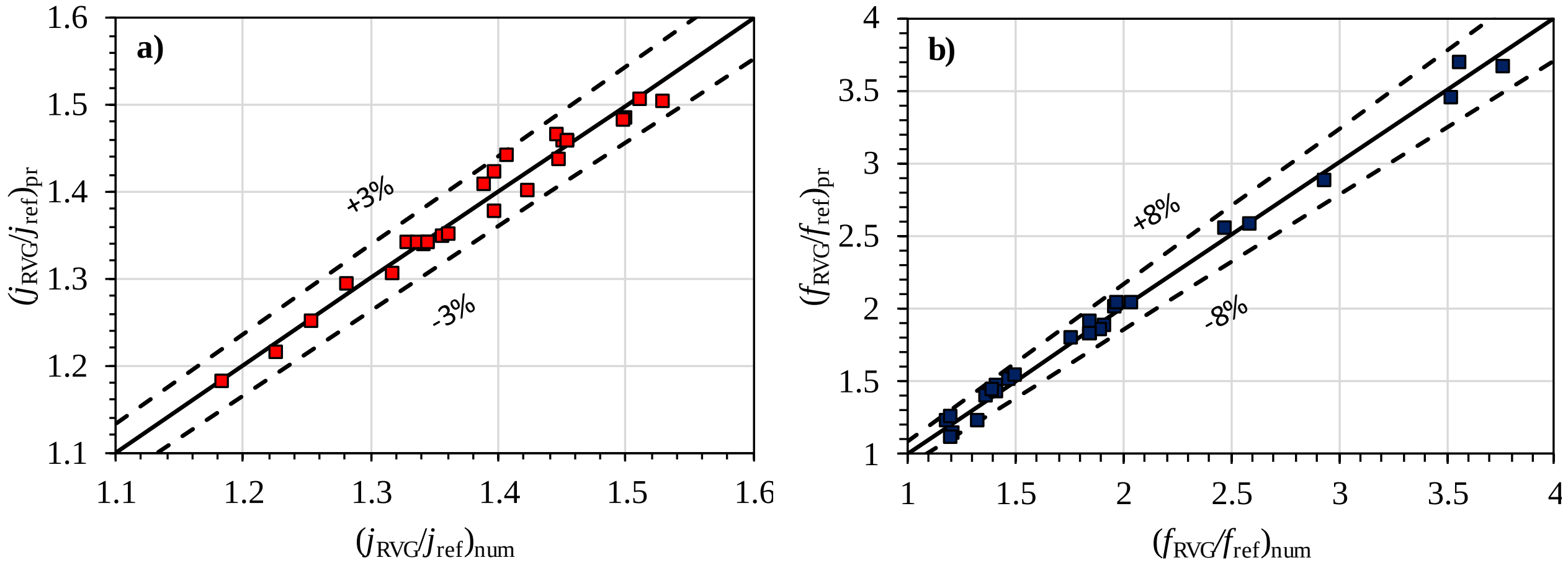

**Fig. 9.** The comparison between predicted and numerically calculated values: a) $f_1 = j_{RVG}/j_{ref}$; b) $f_2 = f_{RVG}/f_{ref}$.

Once the nonlinear regression models are obtained along with their respective parameters, optimization is carried out using the genetic algorithm approach. A single-objective genetic algorithm was utilized to optimize the objective function $F_1$ (Eq. 18), focusing on finding the best solution with respect to a single criterion, i.e. maximizing FTHE heat transfer performance. In contrast, the multi-objective optimization algorithm NSGA-II was used to optimize multiple conflicting objectives simultaneously, i.e. maximizing the heat transfer and minimizing the pressure drop.

The flowchart shown in Fig. 10 outlines a structured optimization process integrating numerical simulation, regression modeling, and application of genetic algorithm to determine optimal RVG geometry.

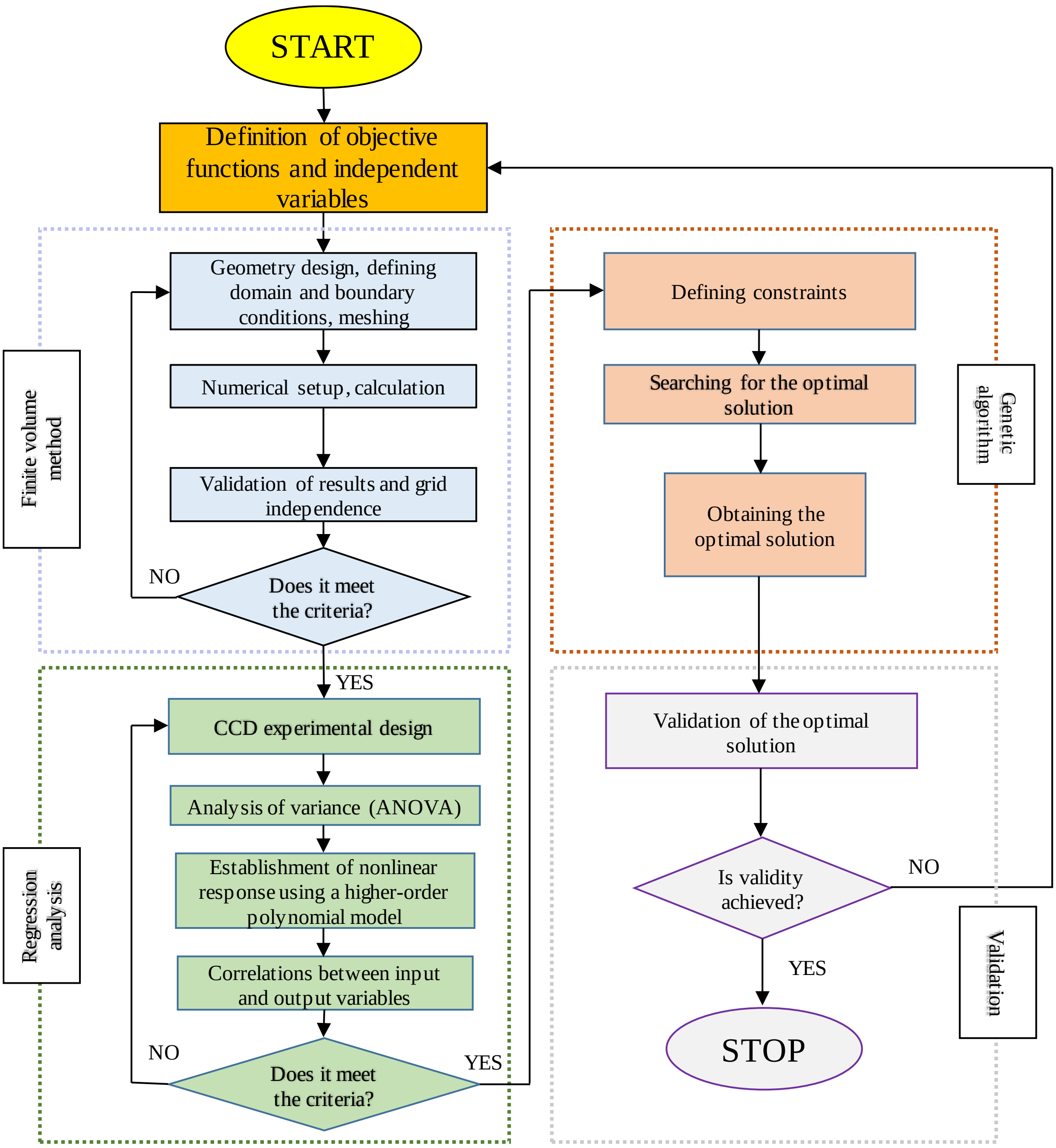

**Fig. 10.** The flow chart of the optimization procedure.

## 5. Optimization results and discussion

### *5.1. Optimization with the objective of heat transfer enhancement*

For the objective function aimed at enhancing heat transfer, the response is defined as the ratio of the Colburn factors of the FTHE with curved winglets and the reference heat exchanger with plain fins ($F_1 = f_1 = j_{RVG}/j_{ref}$). Optimum values of four RVG geometry parameters for the analyzed objective function $F_1$ are shown in Table 5. For the optimized configuration, the enhancement of the Colburn factor is 53.1% in comparison to the reference heat exchanger.

**Table 5** Optimum values of RVG geometry parameters for the objective function $F_1$.

| Geometry parameter | Optimum value |
|---|---|
| Axial position in the airflow $x$ direction, $\Delta X_{\mathrm{RVG}}$ (mm) | 17.73 |
| RVG attack angle, $\alpha_{\mathrm{RVG}}$ (°) | 72.76 |
| RVG height, $H_{\mathrm{RVG}}$ (mm) | 2.03 |
| RVG curvature angle, $\Theta_{\mathrm{RVG}}$ (°) | 32.18 |
| **Objective function** | |
| $F_1 = j_{\mathrm{RVG}}/j_{\mathrm{ref}}$ | 1.531 |

Reliability of the optimum solution obtained using the genetic algorithm is evaluated through comparison with numerical results for the values of geometry parameters listed in Table 5. As seen from Table 6, a very good agreement can be observed between the results obtained with the genetic algorithm and the results based on numerical calculation, with a relative error of 0.326%.

**Table 6** Comparison of numerical results and the results obtained using the genetic algorithm for the objective function $F_1$.

| Method | $F_1 = j_{\mathrm{RVG}}/j_{\mathrm{ref}}$ |
|---|---|
| Genetic algorithm | 1.531 |
| Numerical calculation | 1.536 |
| **Relative error (%)** | 0.326 |

*5.2. Optimization with the objective of enhancing the heat transfer while considering the influence of pressure drop*

Optimization with the objective of enhancing the heat transfer while considering the influence of pressure drop is based on introducing the direct comparability of two solutions. For the considered FTHE with curved RVGs, the objective is to maximize the Colburn factor and minimize the friction factor. Such criterion is often conflicting, since heat transfer enhancement is generally related to larger flow resistance and consequently, larger pressure drops. For such problems, it is necessary to offer more possible optimal solutions, i.e. define the global Pareto-optimal solution set. To determine the Pareto-optimal solution set, two functions, $f_1 = j_{\mathrm{RVG}}/j_{\mathrm{ref}}$ and $f_2 = f_{\mathrm{RVG}}/f_{\mathrm{ref}}$, are contrasted. Obtained set of optimum solutions is shown with the Pareto front in Fig. 11.

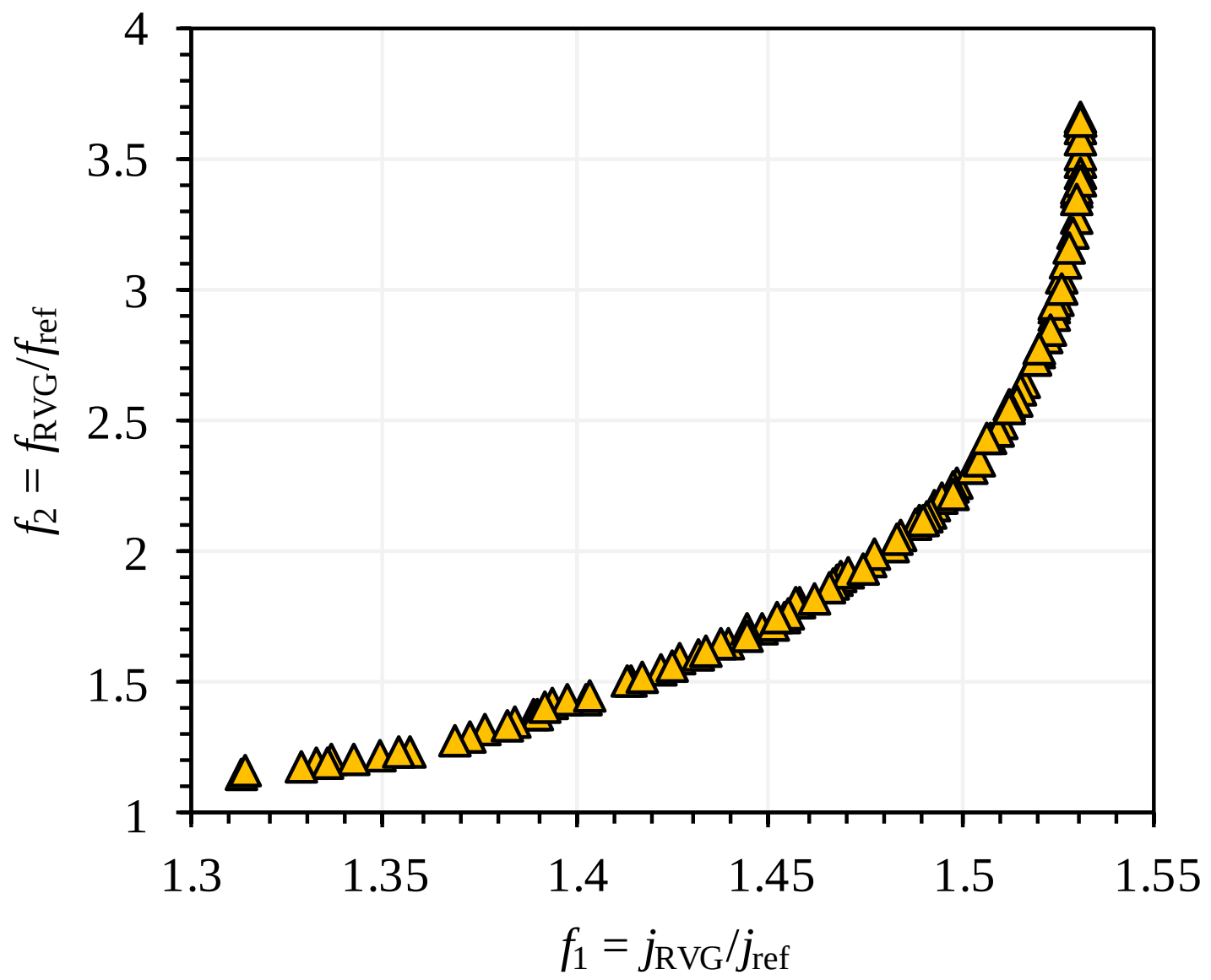

**Fig. 11.** Pareto-optimal solution set.

The objective is to find suitable combinations of RVG geometry parameters for which the values of objective functions $F_2 = f_1/f_2$ and $F_3 = f_1/f_2^{0.33}$ are the largest. The objective function $F_2$ represents the area goodness factor ratio $JF_{\mathrm{I}}$, while the objective function $F_3$ refers to the volume goodness factor ratio $JF_{\mathrm{II}}$. In order to select the optimum solution, a comparison of Pareto solution sets for the objective functions $F_2 = JF_{\mathrm{I}}$ and $F_3 = JF_{\mathrm{II}}$ is shown in Fig. 12.

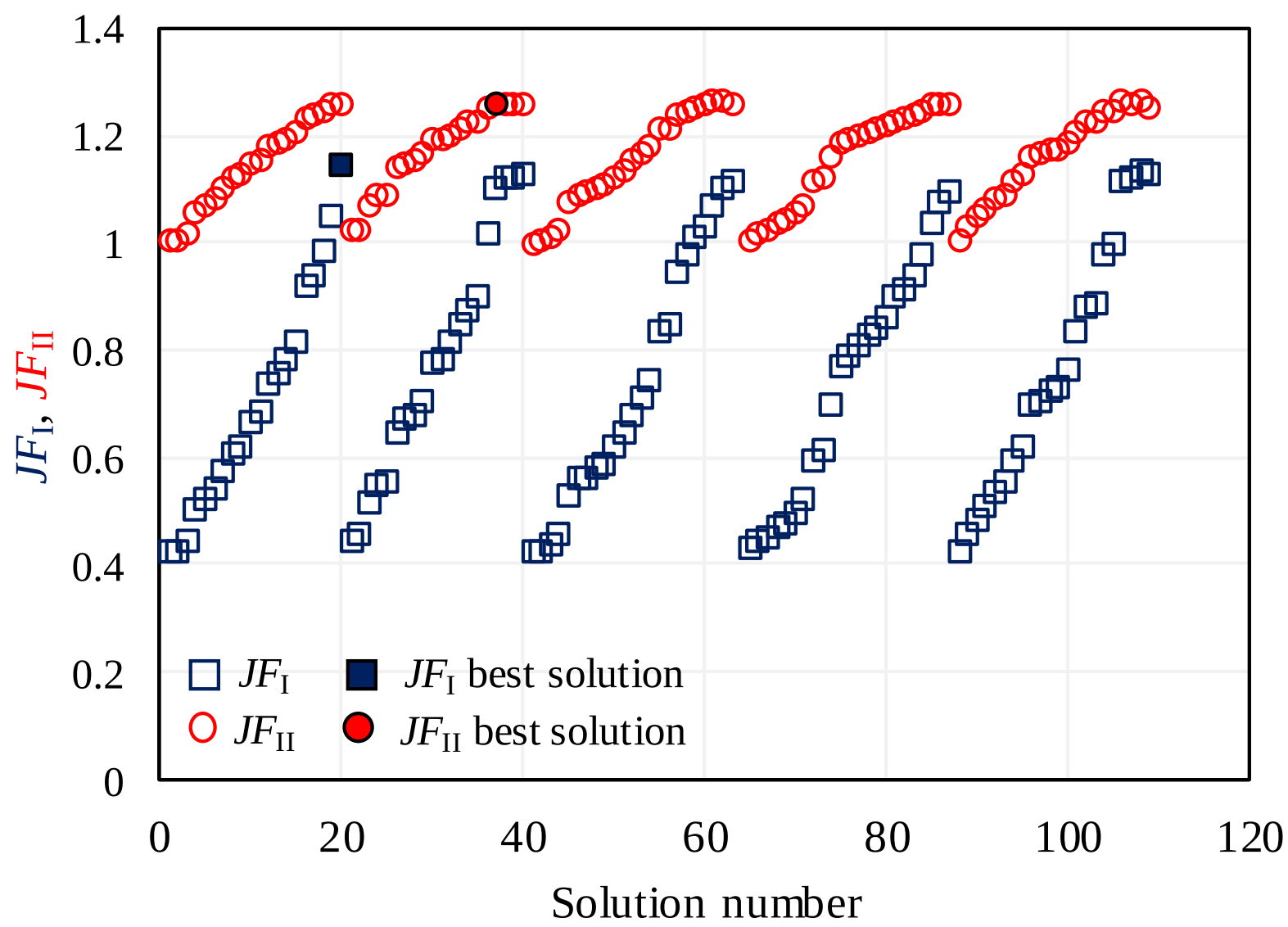

**Fig. 12.** Comparison of factors $JF_{\mathrm{I}}$ and $JF_{\mathrm{II}}$ for the obtained solution set from the Pareto front.

The comparison of the proposed solution set shows that the largest $JF_{\mathrm{I}}$ and $JF_{\mathrm{II}}$ factors are obtained for solutions 20 and 37, respectively, so these solutions are identified as the best from the set of all solutions. The values of analyzed geometry parameters for the optimum solutions are shown in Table 7. For the FTHE with optimized RVG geometry parameters, the area goodness factor $j/f$ is enhanced by 14.4% and the volume goodness factor $j/f^{\,0.33}$ by 26.5% compared to the reference plain FTHE.

**Table 7** Optimum values of RVG geometry parameters for the objective functions $F_2$ and $F_3$ with the aim of enhancing the heat transfer while considering the influence of pressure drop.

| **Geometry parameter** | **Optimum value** | |
|---|---|---|
| | $\boldsymbol{F_2}$ | $\boldsymbol{F_3}$ |
| RVG attack angle, $\alpha_{RVG}$ (°) | 15.39 | 16.21 |
| RVG curvature angle, $\Theta_{RVG}$ (°) | 83.01 | 80.46 |
| Axial position in the airflow $x$ direction, $\Delta X_{RVG}$ (mm) | 17.13 | 13.94 |
| RVG height, $H_{RVG}$ (mm) | 1.83 | 1.70 |
| **Objective function** | | |
| $F_2 = (j_{RVG}/j_{ref})/(f_{RVG}/f_{ref})$ | 1.144 | |
| $F_3 = (j_{RVG}/j_{ref})/\ (f_{RVG}/f_{ref})^{0.33}$ | 1.265 | |

The reliability of the optimum solution obtained with the genetic algorithm is evaluated through comparison with numerical results for the values of geometry parameters presented in Table 7. Table 8 shows good agreement between results obtained with the genetic algorithm and the results of numerical calculations for the functions $f_1$ and $f_2$ as well as for the objective functions $F_2 = f_1/f_2$ and $F_3 = f_1/f_2^{0.33}$. The discrepancies in the $f_2$ values are higher compared to the $f_1$ values, as can be seen from the relative errors, which reach up to about 4%, while the corresponding errors for $f_1$ remain below 0.5%. However, the overall correspondence between the results demonstrates the reliability and robustness of the genetic algorithm for optimizing functions under study.

**Table 8** Comparison of numerical results and the results obtained using the genetic algorithm for the objective functions $F_2$ (a) and $F_3$ (b).

| | (a) | | | (b) | | |
|---|---|---|---|---|---|---|
| Method | $\boldsymbol{f_1}$ | $\boldsymbol{f_2}$ | $\boldsymbol{F_2 = f_1 / f_2}$ | $\boldsymbol{f_1}$ | $\boldsymbol{f_2}$ | $\boldsymbol{F_3 = f_1 / f_2^{0.33}}$ |
| Genetic algorithm | 1.313 | 1.148 | 1.144 | 1.356 | 1.235 | 1.265 |
| Numerical calculation | 1.316 | 1.194 | 1.102 | 1.362 | 1.265 | 1.260 |
| **Relative error (%)** | 0.20 | 4.05 | 3.67 | 0.48 | 2.43 | 0.39 |

*5.3. Performance evaluation of the optimized configurations*

To evaluate the performance of optimized configurations relative to the conventional winglet design, three optimized curved RVG geometries are compared with the corresponding flat-RVG configurations, which are obtained by setting the curvature angle to $\Theta_{RVG} = 0°$, while keeping the other geometry parameters unchanged at their respective values given in Tables 5 and 7. The values of the objective functions $F_1$, $F_2$ and $F_3$ for both sets of configurations are shown in Fig. 13. Although both sets provide improved performance relative to the plain-fin design, the optimized curved configurations achieve higher values for all three objective functions. In comparison with the flat-RVG designs, the increase in $F_1$ is relatively modest (about 3%), whereas the improvements become more pronounced for the thermo-hydraulic criteria, reaching approximately 6.9% for $F_2$ and 9.7% for $F_3$.

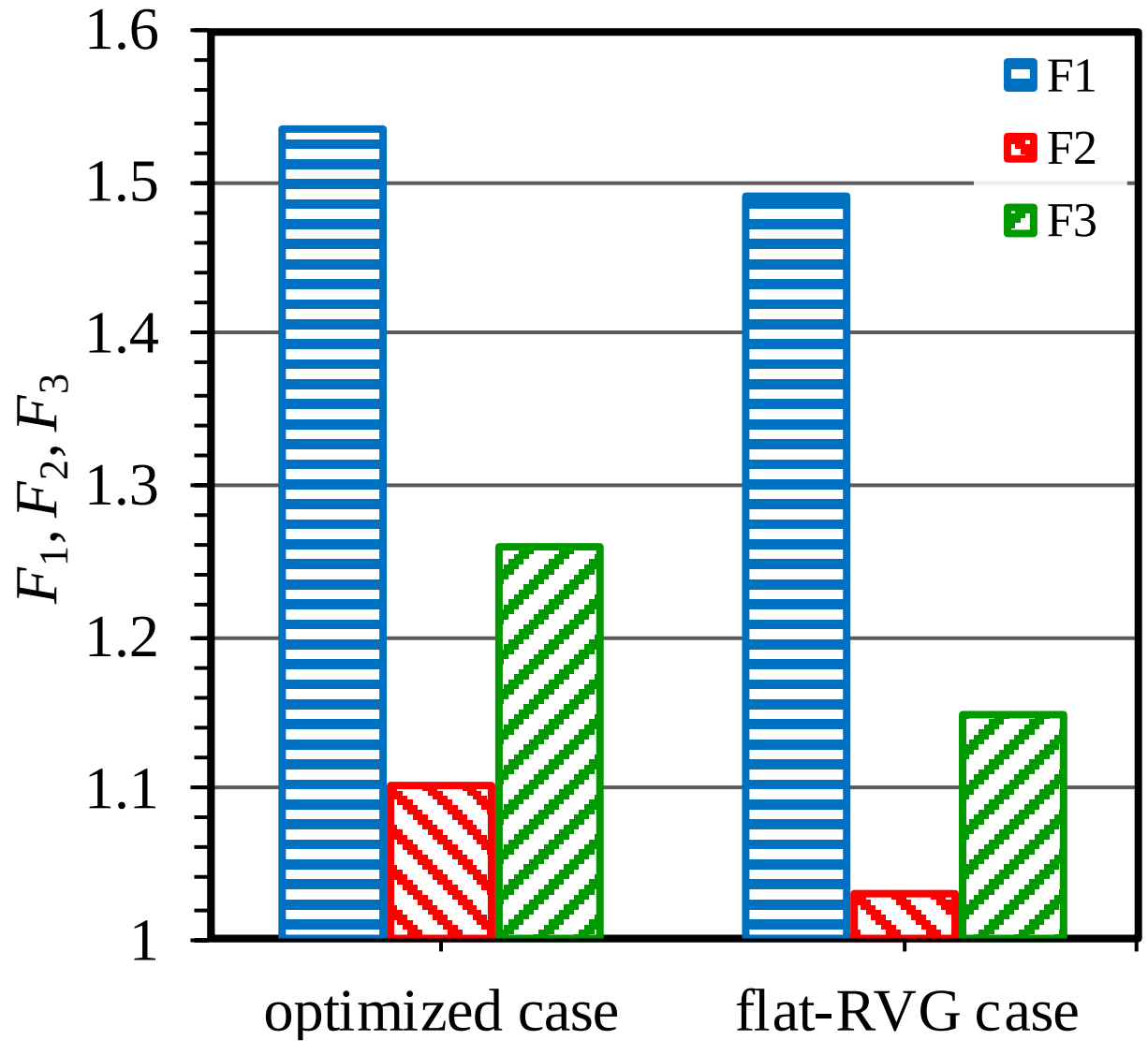

**Fig. 13.** Comparison of performance metrics for the optimized and flat-RVG configurations.

A further comparison was performed numerically by using the CCD central state ($\alpha_{RVG}$ = 45°, $\Theta_{RVG}$ = 45°, $\Delta X_{RVG}$ = 15 mm and $H_{RVG}$ = 1.57 mm) as the reference geometry. For this purpose, the objective functions $F_1$, $F_2$ and $F_3$ were redefined as ratios relative to the central state rather than the plain-fin model. The corresponding objective function values are shown in Table 9. Compared with the new baseline case, the optimized RVG configurations achieve increases of 5.7%, 40.4% and 6.4% for $F_1$, $F_2$ and $F_3$, respectively. These results confirm that the optimized configurations provide improved heat transfer and thermo-hydraulic performance not only when compared with the plain-fin FTHE and the flat-RVG alternatives, but also relative to the representative non-optimized RVG geometry.

**Table 9** Comparison of optimized configurations with the CCD central-point baseline.

| **Geometry parameter** | **Optimum value** | | |
|---|---|---|---|
| | $F_1$ | $F_2$ | $F_3$ |
| RVG attack angle, $\alpha_{RVG}$ (°) | 72.76 | 15.39 | 16.21 |
| RVG curvature angle, $\Theta_{RVG}$ (°) | 32.18 | 83.01 | 80.46 |
| Axial position in the airflow *x* direction, $\Delta X_{RVG}$ (mm) | 17.73 | 17.13 | 13.94 |
| RVG height, $H_{RVG}$ (mm) | 2.03 | 1.83 | 1.70 |
| **Objective function** | | | |
| $^*F_1 = j_{RVG}/j_{ref}$ | 1.057 | | |
| $^*F_2 = (j_{RVG}/j_{ref})/(f_{RVG}/f_{ref})$ | 1.404 | | |
| $^*F_3 = (j_{RVG}/j_{ref})/(f_{RVG}/f_{ref})^{0.33}$ | 1.064 | | |

* The CCD central state is used for the reference configuration

*5.4. Thermal and flow field visualization*

To better understand the effects of optimized RVG geometry parameters on heat transfer and fluid flow inside the FTHE, the numerically obtained temperature contours and velocity vectors in the *xy* midplane ($z = S_f/2$) are shown in Fig. 14. Due to the large attack angle and moderate curvature angle in the optimized $F_1$ configuration (case b), the winglets effectively direct part of the air into the downstream area behind the tube. With the optimized configuration (b), the recirculation zones, which are characterized by low velocity and high temperature, are significantly reduced, which contributes to a more intensive heat transfer. Furthermore, regions of high air velocities can be observed near the second and third tube rows, resulting in greater pressure gradients along the tubes' outer surfaces and formation of more intense horseshoe vortices, which further enhances heat transfer. The optimum RVG parameters found for objective functions $F_2$ (c) and $F_3$ (d) similarly improve the flow mixing in the wake zones behind the tubes, compared to the reference case a) without VGs. When RVGs are configured at a higher attack angle, i.e. at $\alpha_{RVG}$ = 72.76° for the optimum case b), the deflection of mainstream is more aggressively present, resulting in a stronger separation and larger vortex structures. Lower local flow velocities can be seen behind the winglets, as a large part of the kinetic energy is diverted into the vortex formation. In contrast, when RVGs are positioned at a lower attack angle ($\alpha_{RVG}$ = 15.39° and $\alpha_{RVG}$ = 16.21° for optimum cases c) and d), respectively), the deflection is relatively moderate, leading to a more uniform velocity distribution. The temperature contours in the optimized configurations b) – d) clearly demonstrate an improved heat transfer performance, with configuration b) showing a significant reduction in the high-temperature regions downstream of the tube surfaces.

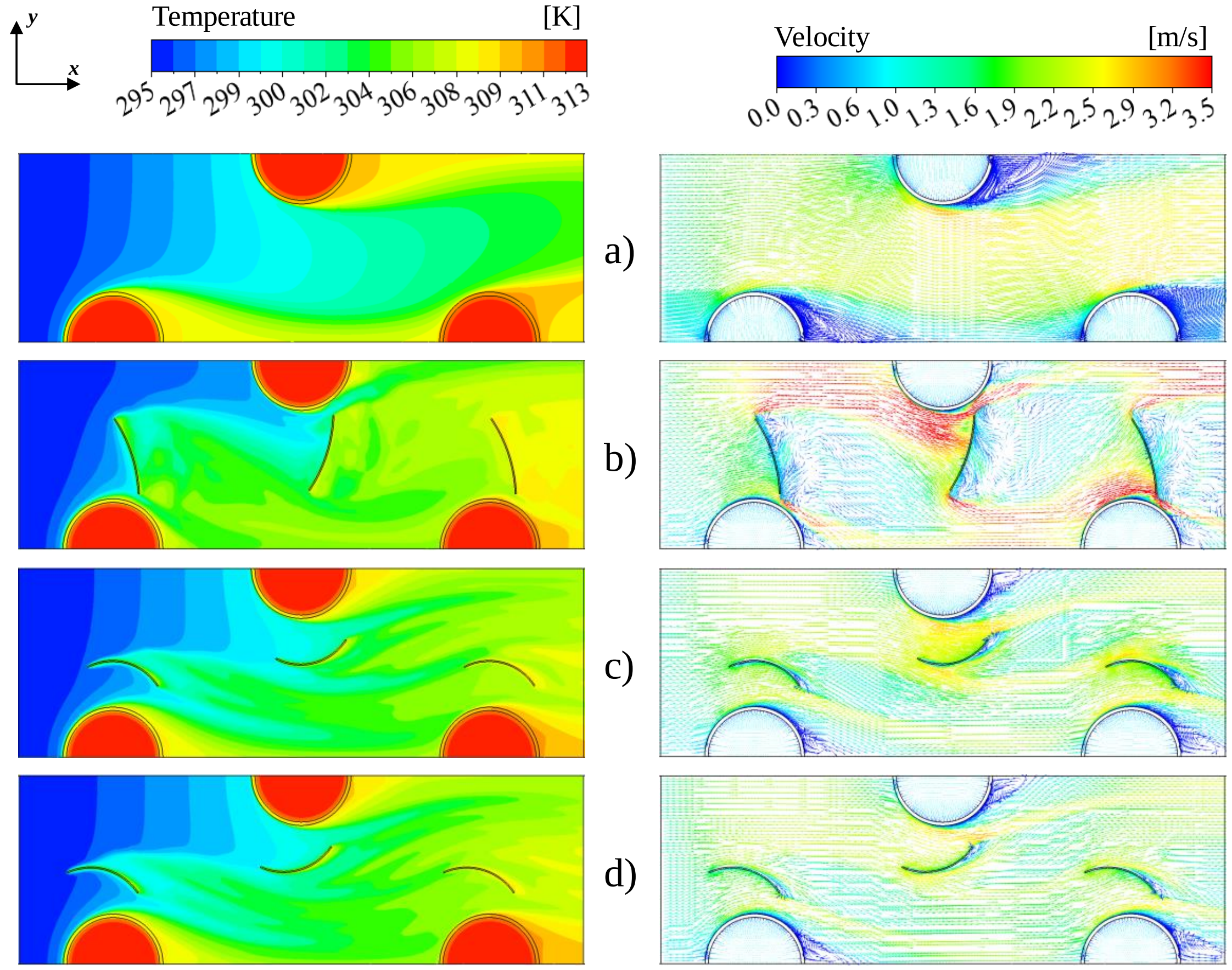

**Fig. 14.** Comparison of temperature contours and velocity vectors in the middle *xy* plane ($z = S_f/2$) for the reference plain FTHE a), the configuration optimized according to the objective of heat transfer enhancement ($F_1$ model) b) and the configurations optimized according to the objective of heat transfer enhancement, while considering the influence of pressure drop using $F_2 = f_1/f_2$ c) and $F_3 = f_1/f_2^{0.33}$ d).

The inclusion of vortex generators alters the local airflow by inducing swirling motions that intensify mixing and disrupt the boundary layer. The resulting complex flow patterns contribute to a more uniform temperature distribution of the air inside the FTHE. The local secondary flow strength $Se_{local}$ [31] was analyzed at several cross-sections (*yz* plane) along airflow direction (Fig. 15), i.e. at $x$ = 20 mm, $x$ = 30 mm, $x$ = 60 mm and $x$ = 80 mm, for both reference and optimized FTHE configurations. The contours of $Se_{local}$ show not only the intensity of the longitudinal vortices generated by optimized winglets, but also the vortex structure, which consists of main, corner and induced vortices.

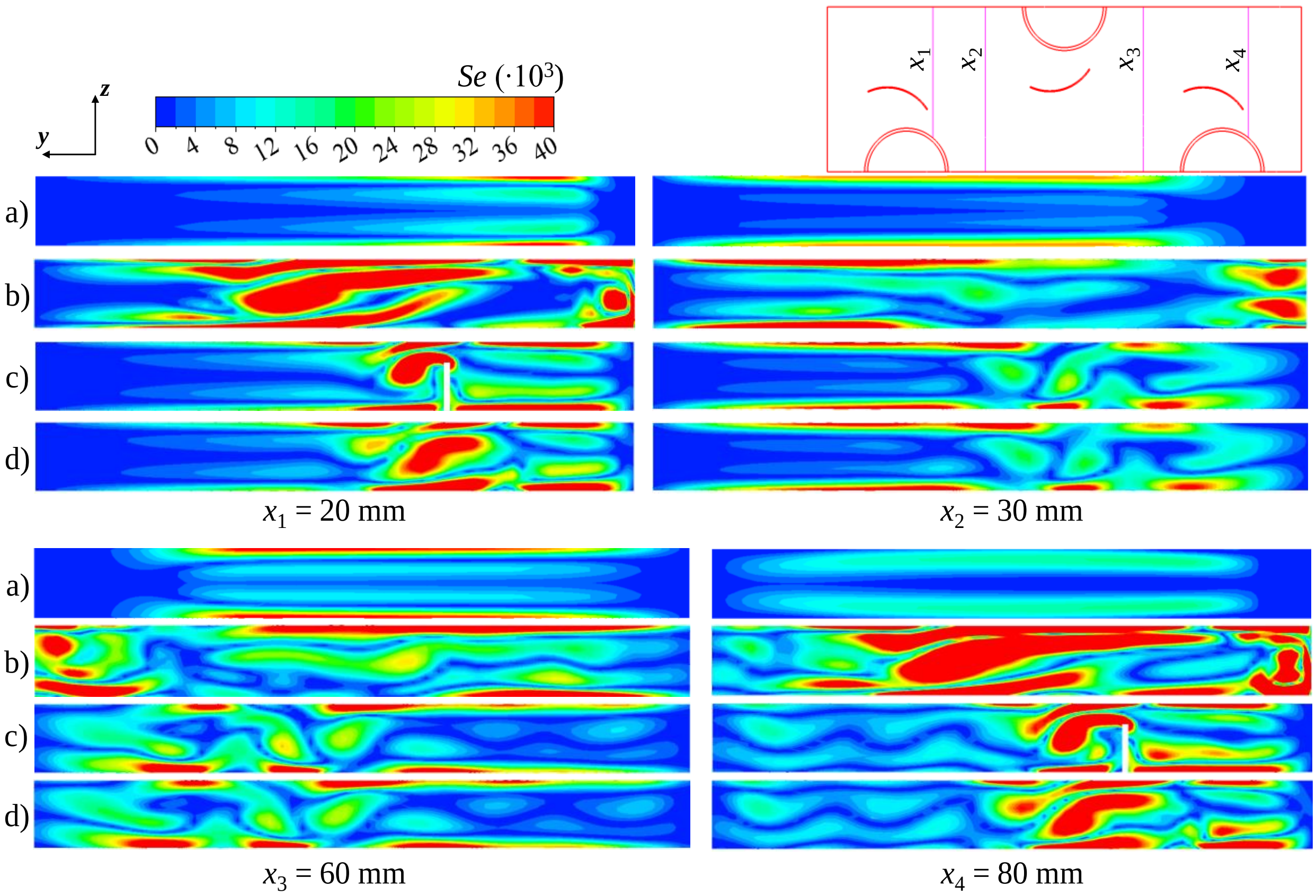

**Fig. 15.** Comparison of local secondary strength distributions for the reference plain FTHE without VGs a), the configuration optimized according to the objective of heat transfer enhancement ($F_1$ model) b) and the configurations optimized according to the objective of heat transfer enhancement, while considering the influence of pressure drop using $F_2 = f_1/f_2$ c) and $F_3 = f_1/f_2^{0.33}$ d), at four streamwise positions for $Re_a$ = 1524.

The main vortices generated by the optimized RVG geometries for the objective functions $F_2$ (c) and $F_3$ (d) are much smaller than those generated by the optimized winglet geometry for the objective function $F_1$ (b), as can be observed at streamwise positions $x$ = 20 mm and $x$ = 80 mm. This is to be expected given the larger attack angle and the greater height in the configuration b). In addition, due to the higher velocities at which the air impinges on the front surface of the tube, stronger horseshoe vortices are formed, which maintain the magnitude at a greater distance from the first and second tube, as can be seen at the positions $x$ = 30 mm and $x$ = 60 mm. The local *Se* distributions show that the longitudinal vortices augment the convective heat transfer on both lower and upper fins, as the optimized RVG heights for the objective functions considered are comparable to the fin spacing, i.e. with greater winglet height, vortex motion transports more kinetic energy into the boundary layer that forms along the surface of the fin where no winglets are mounted (upper fin). The increased secondary flow interferes with the boundary layer, reducing its thickness and improving heat transfer. This consequently leads to higher air temperatures at the heat exchanger's outlet, as shown in Fig. 16 for the *xz* midplane ($y$ = 15 mm). In the reference configuration without winglet inserts, the thermal boundary layer gradually forms along the airflow direction. Optimized curved RVGs create localized turbulent structures that entrain cooler air into the near-wall region, promoting boundary layer restoration and thereby reducing the air-side convective thermal resistance.

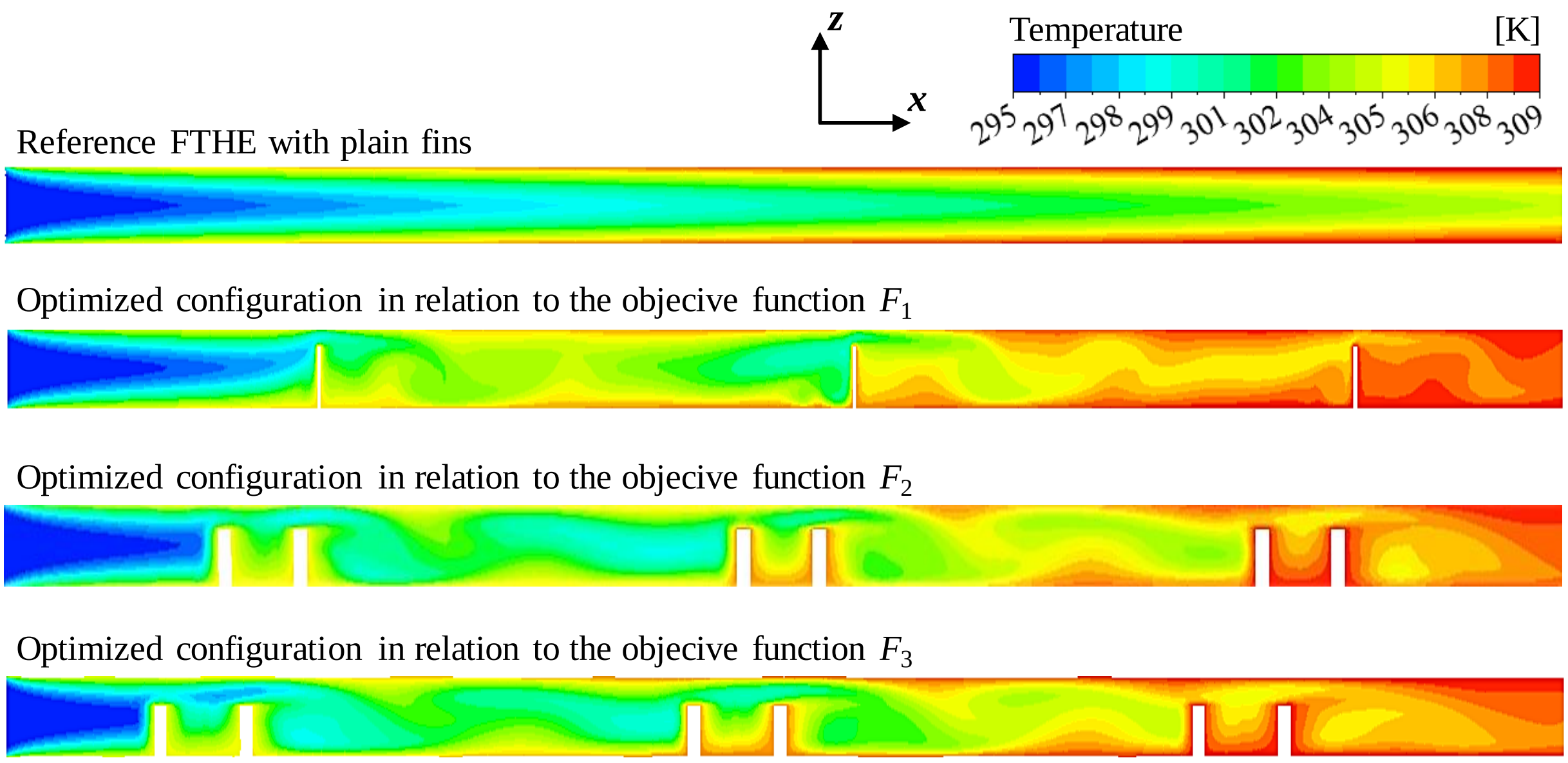

**Fig. 16.** Comparison of temperature contours on the air side at *xz* midplane ($y$ = 15 mm) for reference and optimized FTHE configurations.

In addition, the contours of local air-side heat transfer coefficient (HTC) on the lower (with RVGs) and upper (without RVGs) fin surfaces have been compared and shown in Fig. 17. The HTC at each point of the fin surface is calculated according to [32].

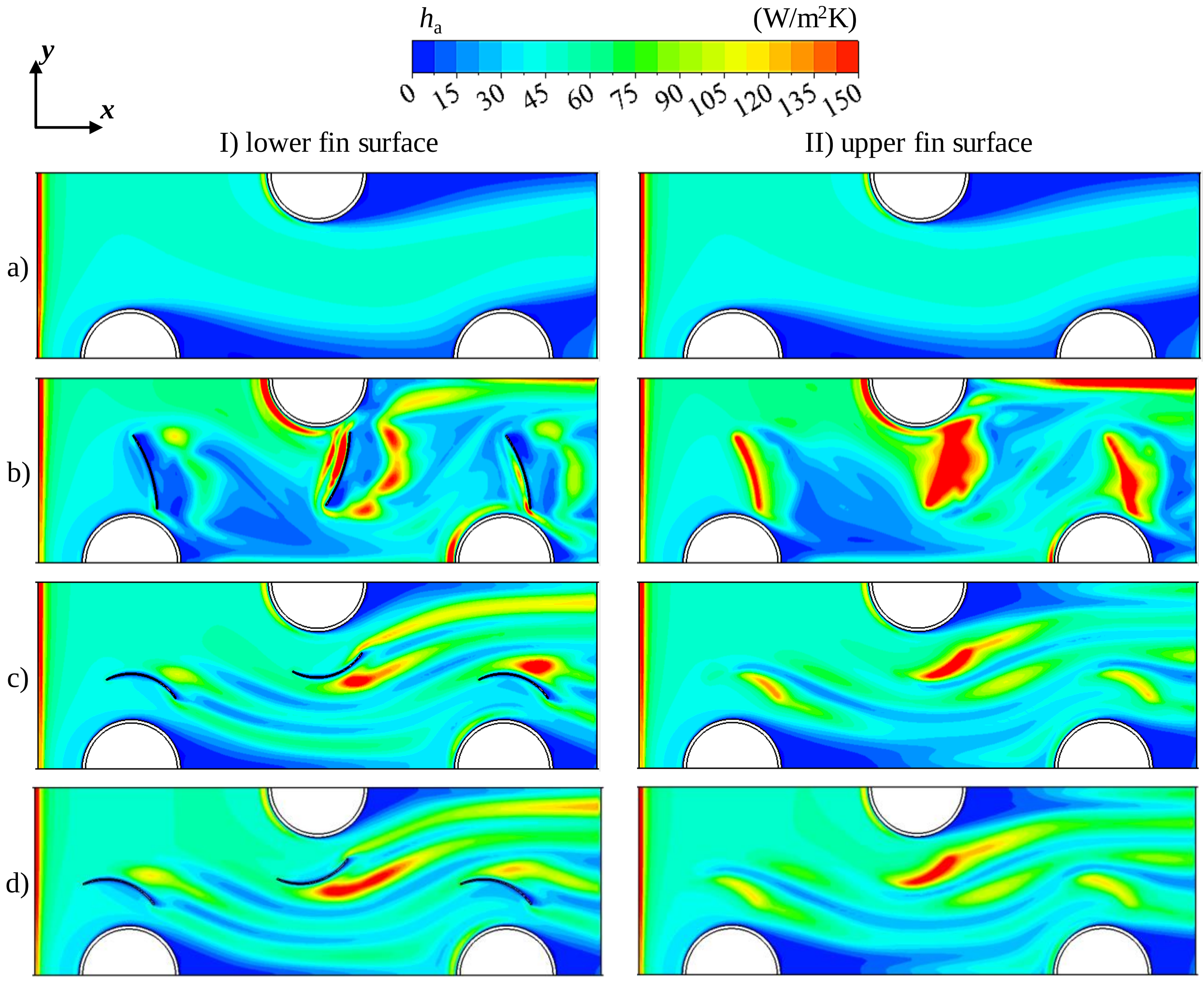

**Fig. 17.** Comparison of local air-side HTC distributions for the reference plain FTHE without VGs a), the configuration optimized according to the objective of heat transfer enhancement ($F_1$ model) b) and the configurations optimized according to the objective of heat transfer enhancement, while considering the influence of pressure drop using $F_2 = f_1/f_2$ c) and $F_3 = f_1/f_2^{0.33}$ d), on I) lower and II) upper fin surfaces; $Re_a = 1524$.

As shown in Fig. 17, the HTC is high near the entrance regions for cases a) – d) as a result of the thin boundary layer and large temperature gradients between the air and the fin surfaces. The HTC gradually decreases as the boundary layer develops along the airflow direction. For the reference configuration a), value of HTC is very small on both sides of the fin surface in the areas behind each tube, due to lower velocities and poor mixing associated with flow separation. When using optimized RVGs, the boundary layer separation is delayed due to the acceleration of air in the narrowed passage between the winglet and the tube. This leads to a reduction in wake size and consequently results in an increase of the HTC, especially in case b). For the optimized configuration b), high HTC values are distributed behind the winglets, in the zones of the formed longitudinal vortices (lower fin surface), and in the zones above the RVGs, attributed to higher local velocities (upper fin surface). For the optimized configurations c) and d), the high HTC zones are distributed near the leeward side of RVGs (lower fin surface) and in the zones above the RVGs (upper fin surface). It can also be concluded that the winglets placed next to the second tube row contribute most to the enhancement of heat transfer.

*5.5. Off-design performance analysis*

Finally, the performance of FTHE with optimized RVG geometry parameters under off-design test conditions (at varying $Re_a$) is assessed. The numerical simulations were performed for $Re_a$ = 2220 and $Re_a$ = 2960, corresponding to air inlet velocities of $w_{a,in}$ = 1.5 m/s and $w_{a,in}$ = 2 m/s, respectively. The results have been shown in Fig. 18. For the FTHE with RVG geometry optimized according to the objective of heat transfer enhancement (optimum values: $\alpha_{RVG}$ = 72.76°, $\Theta_{RVG}$ = 32.18°, $\Delta X_{RVG}$ = 17.73 mm, $H_{RVG}$ = 2.03 mm), the Colburn factor increases by 53.6%, 76.1% and 91.9% for $Re_a$ = 1524, 2220 and 2960, respectively, compared to the reference FTHE with plain fins. However, the aforementioned configuration is unfavorable from the aspect of combined thermo-hydraulic performance of the FTHE, especially when equal importance is given to heat transfer and pressure drop, due to high flow resistance caused by the large RVG attack angle. If the objective is to increase heat transfer, while taking into account the influence of pressure drop through the objective function $F_2 = JF_I$ (Eq. 19), the FTHE with the optimum RVG geometry parameters ($\alpha_{RVG}$ = 15.39°, $\Theta_{RVG}$ = 83.01°, $\Delta X_{RVG}$ = 17.13 mm, $H_{RVG}$ = 1.83 mm) achieves an enhancement of the $j/f$ factor by 10.2%, 11.3% and 15.1% for $Re_a$ = 1524, 2220 and 2960, respectively, compared to the reference plain-fin model. For the objective function $F_3 = JF_{II}$ (Eq. 20), the optimum set of values ($\alpha_{RVG}$ = 16.21°, $\Theta_{RVG}$ = 80.46°, $\Delta X_{RVG}$ = 13.94 mm, $H_{RVG}$ = 1.7 mm) results in an improvement of the factor $j/f^{0.33}$ by 26%, 35.1% and 40.9% compared to the reference case, for the $Re_a$ considered.

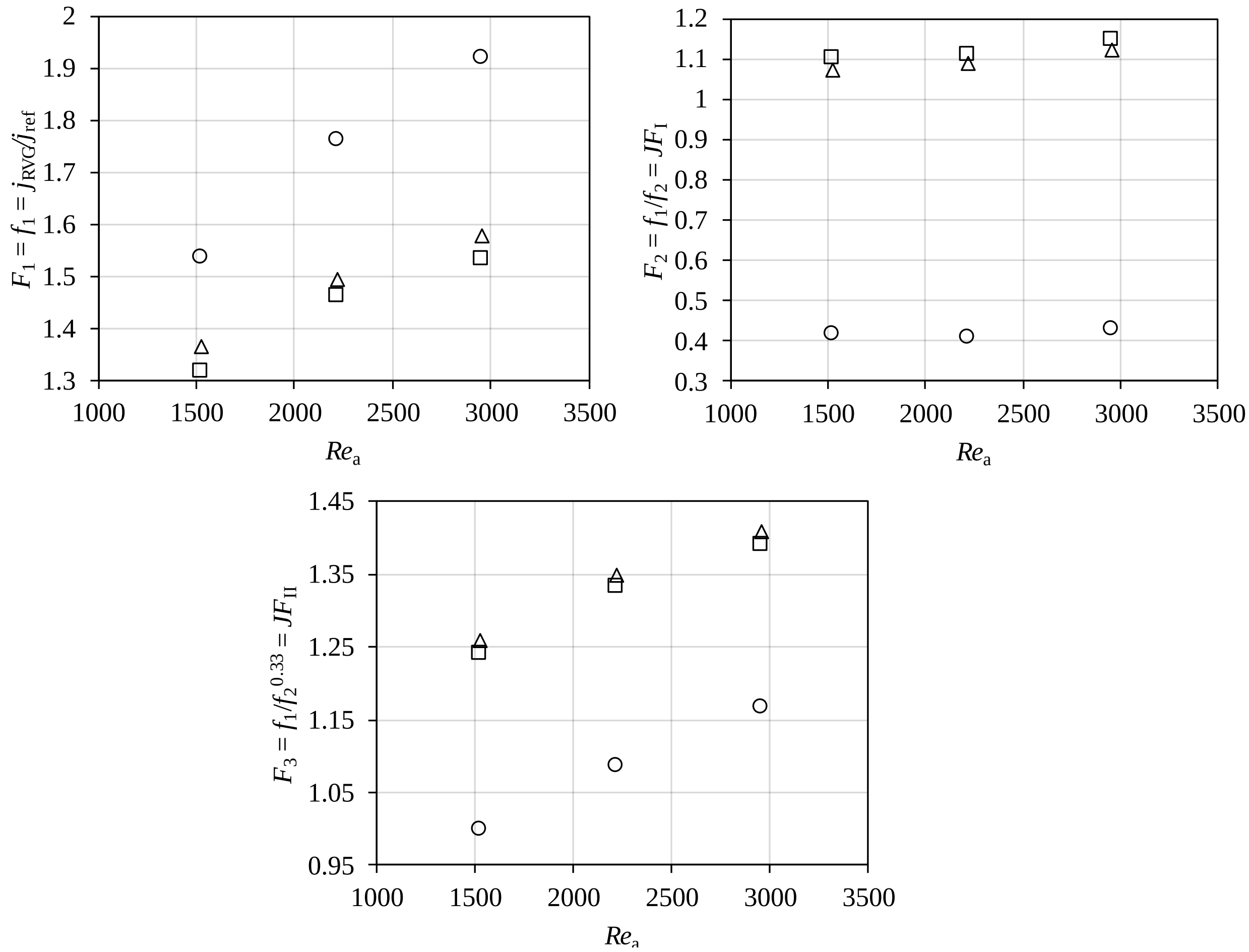

○ - optimized $F_1$ RVG configuration, □ - optimized $F_2$ RVG configuration,

△ - optimized $F_3$ RVG configuration

**Fig. 18.** A performance comparison of the optimized configurations for $Re_a$ = 1524 as well as for off-design test conditions ($Re_a$ = 2220 and $Re_a$ = 2960), obtained numerically.

## 6. Conclusion

In the present numerical investigation, the curved rectangular vortex generators have been installed on fin surfaces inside water-to-air finned-tube heat exchanger. Four geometry parameters ($\alpha_{RVG}$, $\Theta_{RVG}$, $\Delta X_{RVG}$ and $H_{RVG}$) have been optimized with respect to two criteria: maximization of heat transfer solely, through objective function $F_1 = j_{RVG}/j_{ref}$, and maximization of heat transfer while considering the influence of pressure drop, through objective functions $F_2 = (j_{RVG}/j_{ref})/(f_{RVG}/f_{ref})$ and $F_3 = (j_{RVG}/j_{ref})/(f_{RVG}/f_{ref})^{0.33}$. The basis of the conducted optimization has been a validated steady-state 3D numerical model applied to the problem of fluid flow and heat transfer, which takes into consideration a fully developed water flow in the tubes. The regression analysis has been used and the Central composite design of experiment was selected in order to obtain non-linear regression polynomial functions $f_1 = j_{RVG}/j_{ref}$ and $f_2 = f_{RVG}/f_{ref}$, which were integrated into the investigated objectives. Once the nonlinear regression models were acquired along with their respective parameters, optimization has been carried out using the genetic algorithm. Reliability of the optimum solution obtained by means of the genetic algorithm was evaluated through comparison with numerical results for the values of optimum geometry parameters and very good agreement between the two has been observed. The main findings are:

- For the objective of maximizing the convective heat transfer on the air side, the optimum combination of analyzed geometry parameters is $\alpha_{\text{RVG}}$ = 72.76°, $\Theta_{\text{RVG}}$ = 32.18°, $\Delta X_{\text{RVG}}$ = 17.73 mm and $H_{\text{RVG}}$ = 2.03 mm, resulting in an enhancement of the Colburn factor by 53.1% when compared with the reference heat exchanger configuration with plain fins.

- For the objective of maximizing the heat transfer while considering the influence of pressure drop through the objective function $F_2$, the optimum combination of analyzed geometry parameters is $\alpha_{\text{RVG}}$ = 15.39°, $\Theta_{\text{RVG}}$ = 83.01°, $\Delta X_{\text{RVG}}$ = 17.13 mm, $H_{\text{RVG}}$ = 1.83 mm, resulting in an enhancement of the performance factor $j/f$ by 14.4% when compared with the reference heat exchanger configuration with plain fins.

- For the objective of maximizing the heat transfer while considering the influence of pressure drop through the objective function $F_3$, the optimum combination of analyzed geometry parameters is $\alpha_{\text{RVG}}$ = 16.21°, $\Theta_{\text{RVG}}$ = 80.46°, $\Delta X_{\text{RVG}}$ = 13.94 mm, $H_{\text{RVG}}$ = 1.7 mm, resulting in an enhancement of the thermal performance factor $j/f^{\,0.33}$ by 26.5% when compared with the reference heat exchanger configuration with plain fins.

- Relative to the corresponding flat-RVG designs ($\Theta_{\text{RVG}}$ = 0°), the optimized curved-RVG configurations achieve increases of 3%, 6.9% and 9.7% for $F_1$, $F_2$ and $F_3$, respectively.

- Under off-design conditions ($Re_{\text{a}}$ = 2220 and $Re_{\text{a}}$ = 2960), the numerical model optimized for the objective function $F_1$ shows a significant heat transfer enhancement, with an increase in the Colburn factor of up to 91.9%, while the configurations optimized according to the objective functions $F_2$ and $F_3$ achieve an improvement in thermo-hydraulic performance of up to 15.1% and 40.9%, respectively, compared to the reference model with plain fins.

The results show that the optimal geometry depends strongly on the selected objective function. The configuration that maximizes heat transfer differs significantly from those obtained when pressure-drop effects are considered, especially regarding the attack angle and curvature angle. This highlights the importance of treating the attack angle as an independent design parameter in the optimization of curved VGs. In addition, the results for off-design conditions indicate that the optimized configurations retain improved performance beyond the specified operating conditions. The optimization results serve as a basis for the development of more efficient FTHEs. Future research will focus on optimization strategies that consider different designs of fins, tubes and vortex generators to further improve the thermal performance of FTHEs.

## Nomenclature

| | |
|---|---|
| $A_f$ | surface area of the fin ($m^2$) |
| $A_{min}$ | minimum flow cross-sectional area ($m^2$) |
| $A_o$ | total outside surface area on the air side ($m^2$) |
| $a$ | thermal diffusivity ($m^2/s$) |
| $a_t$ | turbulent thermal diffusivity ($m^2/s$) |
| $Cd_\omega$ | cross-diffusion term in turbulence model |
| $c_p$ | specific heat capacity at constant pressure (J/kgK) |
| $D$ | tube diameter (m) |
| $D_k$ | dissipation term in *k*-equation |
| $D_\omega$ | dissipation term in $\omega$-equation |
| $F_1, F_2, F_3$ | objective functions |
| $f$ | friction factor |
| $G_k$ | generation term in *k*-equation |
| $G_\omega$ | generation term in $\omega$-equation |
| $H$ | height (m) |
| $h$ | heat transfer coefficient ($W/m^2K$) |
| $j$ | Colburn factor |
| $JF_I$ | area goodness factor ratio |
| $JF_{II}$ | volume goodness factor ratio (thermal performance factor) |
| $k$ | turbulent kinetic energy ($m^2/s^2$) |
| $L_f$ | fin length (m) |
| $L_t$ | tube length (m) |
| $Nu$ | Nusselt number |
| $n_t$ | number of tube rows |
| $P_L$ | streamwise pitch of the tube (m) |
| $P_T$ | crosswise pitch of the tube (m) |
| $Pr$ | Prandtl number |
| $\dot{Q}$ | exchanged heat flux (W) |
| $\dot{q}$ | specific heat flux ($W/m^2$) |
| $R^2$ | coefficient of determination |
| $R_{adj}^2$ | adjusted coefficient of determination |
| $Re$ | Reynolds number |
| $S_f$ | fin pitch (m) |
| $Se$ | dimensionless secondary flow strength |
| $T$ | temperature (K) |
| $t_f$ | thickness of the fin (m) |
| $U$ | overall heat transfer coefficient ($W/m^2K$) |
| $W_f$ | fin width (m) |
| $w$ | velocity (m/s) |
| $w_m$ | mean velocity at minimum flow cross-section (m/s) |

*Greek symbols*

| | |
|---|---|
| $\alpha_{RVG}$ | attack angle of the winglet (°) |
| $\delta_{RVG}$ | thickness of the winglet (m) |
| $\delta_{ij}$ | Kronecker delta |
| $\Delta p$ | pressure drop (Pa) |
| $\Delta T$ | temperature difference (K) |
| $\Delta X_{RVG}$ | winglet axial position in the *x* direction (m) |
| $\Delta Y_{RVG}$ | winglet axial position in the *y* direction (m) |
| $\eta_o$ | extended surface efficiency |
| $\Theta_{RVG}$ | curvature angle of the winglet (°) |
| $\lambda$ | thermal conductivity (W/mK) |
| $\nu$ | kinematic viscosity ($m^2$/s) |
| $\nu_t$ | kinematic turbulent viscosity ($m^2$/s) |
| $\rho$ | density (kg/$m^3$) |
| $\omega$ | turbulent dissipation rate (1/s) |

*Subscripts*

| | |
|---|---|
| a | air |
| b | bulk |
| f | fin |
| i | inside |
| in | inlet |
| lm | logarithmic mean |
| o | outside |
| out | outlet |
| RVG | rectangular vortex generator |
| s | solid wall |
| t | tube |
| w | water |

## CRediT authorship contribution statement

**Josip Batista:** Writing - original draft, Writing - review and editing, Conceptualization, Methodology, Investigation, Data curation, Visualization, Formal analysis, Validation, Software. **Anica Trp:** Writing - review and editing, Conceptualization, Methodology, Visualization, Validation, Supervision, Funding acquisition, Project administration, Resources. **Kristian Lenić:** Writing - review and editing, Conceptualization, Methodology, Visualization, Validation, Supervision, Resources. **Mateo Kirinčić:** Writing - review and editing, Conceptualization, Visualization.

## Acknowledgements

This work has been funded by the European Union – NextGenerationEU – HEEGES – uniri-mzi-25-33 and by Croatian Science Foundation – HEXENER – IP-2016-06-4095.